\documentclass[draftclsnofoot,twoside,onecolumn,letter,12pt]{IEEEtran}
\usepackage{graphicx}
\usepackage{amssymb}
\usepackage{amsmath,cases}
\usepackage{cite}
\usepackage{stfloats}
\usepackage{subfigure}
\usepackage{epstopdf}
\usepackage{mdwlist}
\usepackage{threeparttable}
\usepackage[all,2cell]{xy} \UseAllTwocells
\usepackage{booktabs}
\usepackage{fancyvrb}
\usepackage{rotating}
\usepackage{rotfloat}
\usepackage{pifont}
\usepackage{verbatim}

\usepackage{bm}

\DeclareMathAlphabet{\mathsfbr}{OT1}{cmss}{m}{n}
\SetMathAlphabet{\mathsfbr}{bold}{OT1}{cmss}{bx}{n}
\DeclareRobustCommand{\msf}[1]{%
  \ifcat\noexpand#1\relax\msfgreek{#1}\else\mathsfbr{#1}\fi
}

\makeatletter
\newcommand{\msfgreek}[1]{\csname s\expandafter\@gobble\string#1\endcsname}
\makeatother

\DeclareFontEncoding{LGR}{}{} 
\DeclareSymbolFont{sfgreek}{LGR}{cmss}{m}{n}
\SetSymbolFont{sfgreek}{bold}{LGR}{cmss}{bx}{n}
\DeclareMathSymbol{\salpha}{\mathord}{sfgreek}{`a}
\DeclareMathSymbol{\sbeta}{\mathord}{sfgreek}{`b}
\DeclareMathSymbol{\sgamma}{\mathord}{sfgreek}{`g}
\DeclareMathSymbol{\sdelta}{\mathord}{sfgreek}{`d}
\DeclareMathSymbol{\sepsilon}{\mathord}{sfgreek}{`e}
\DeclareMathSymbol{\szeta}{\mathord}{sfgreek}{`z}
\DeclareMathSymbol{\seta}{\mathord}{sfgreek}{`h}
\DeclareMathSymbol{\stheta}{\mathord}{sfgreek}{`j}
\DeclareMathSymbol{\siota}{\mathord}{sfgreek}{`i}
\DeclareMathSymbol{\skappa}{\mathord}{sfgreek}{`k}
\DeclareMathSymbol{\slambda}{\mathord}{sfgreek}{`l}
\DeclareMathSymbol{\smu}{\mathord}{sfgreek}{`m}
\DeclareMathSymbol{\snu}{\mathord}{sfgreek}{`n}
\DeclareMathSymbol{\sxi}{\mathord}{sfgreek}{`x}
\DeclareMathSymbol{\somicron}{\mathord}{sfgreek}{`o}
\DeclareMathSymbol{\spi}{\mathord}{sfgreek}{`p}
\DeclareMathSymbol{\srho}{\mathord}{sfgreek}{`r}
\DeclareMathSymbol{\ssigma}{\mathord}{sfgreek}{`s}
\DeclareMathSymbol{\stau}{\mathord}{sfgreek}{`t}
\DeclareMathSymbol{\supsilon}{\mathord}{sfgreek}{`u}
\DeclareMathSymbol{\sphi}{\mathord}{sfgreek}{`f}
\DeclareMathSymbol{\schi}{\mathord}{sfgreek}{`q}
\DeclareMathSymbol{\spsi}{\mathord}{sfgreek}{`y}
\DeclareMathSymbol{\somega}{\mathord}{sfgreek}{`w}

\DeclareMathSymbol{\svarsigma}{\mathord}{sfgreek}{`c}

\DeclareMathSymbol{\sGamma}{\mathalpha}{sfgreek}{`G}
\DeclareMathSymbol{\sDelta}{\mathalpha}{sfgreek}{`D}
\DeclareMathSymbol{\sTheta}{\mathalpha}{sfgreek}{`J}
\DeclareMathSymbol{\sLambda}{\mathalpha}{sfgreek}{`L}
\DeclareMathSymbol{\sXi}{\mathalpha}{sfgreek}{`X}
\DeclareMathSymbol{\sPi}{\mathalpha}{sfgreek}{`P}
\DeclareMathSymbol{\sSigma}{\mathalpha}{sfgreek}{`S}
\DeclareMathSymbol{\sUpsilon}{\mathalpha}{sfgreek}{`U}
\DeclareMathSymbol{\sPhi}{\mathalpha}{sfgreek}{`F}
\DeclareMathSymbol{\sPsi}{\mathalpha}{sfgreek}{`Y}
\DeclareMathSymbol{\sOmega}{\mathalpha}{sfgreek}{`W}

\DeclareRobustCommand{\mcal}[1]{%
  \ifcat\noexpand#1\relax\mathnormal{#1}\else\cal{#1}\fi
}
\DeclareRobustCommand{\BM}[1]{%
  \ifcat\noexpand#1\relax\bm{\boldUppercaseItalicGreek{#1}}\else\bm{#1}\fi
}
\makeatletter
\newcommand{\boldUppercaseItalicGreek}[1]{\csname var\expandafter\@gobble\string#1\endcsname}
\makeatother
\newcommand{\rv}[1]{\msf{#1}}
\newcommand{\RV}[1]{\bm{\msf{#1}}}

\let\geq\geqslant

\let\leq\leqslant

\DeclareMathAlphabet{\mathpzc}{OT1}{pzc}{m}{it}

\newcommand{\BB}[1]{\pmb{#1}}
\newcommand{\pDefine}[1]{\mathpzc{#1}}

\newcommand{\oSeq}{\BB{\pDefine{O}}}

\newcommand{\pSeq}{\BB{\pDefine{P}}}
\newcommand{\pC}{\pDefine{c}}
\newcommand{\pM}{\pDefine{m}}
\newcommand{\pN}{\pDefine{n}}
\newcommand{\pP}{\pDefine{p}}
\newcommand{\pQ}{\pDefine{q}}
\newcommand{\pK}{\pDefine{k}}
\newcommand{\pA}[1]{\pDefine{a}_{#1}}
\newcommand{\pB}[1]{\pDefine{b}_{#1}}
\newcommand{\pS}[1]{\pDefine{A}_{#1}}
\newcommand{\pT}[1]{\pDefine{B}_{#1}}
\newcommand{\aV}{\BB{\pDefine{a}}}
\newcommand{\bV}{\BB{\pDefine{b}}}
\newcommand{\sV}{\BB{\pDefine{A}}}
\newcommand{\tV}{\BB{\pDefine{B}}}

\newcommand{\Fox}[5]{H^{#1,#2}_{#3,#4}\left(#5\right)}

\newcommand{\FoxH}[7]{H^{#1,#2}_{#3,#4}
                        \left[
                            {#5}
                            \left|
                            \begin{array}{c}
                                {#6}
                                \\
                                {#7}
                            \end{array}
                            \right.
                        \right]}

\newcommand{\FoxV}[5]{\mathscr{H}^{#1,#2}_{#3,#4}\left(#5\right)}

\newcommand{\stdOP}[2]{#1 \boxdot #2}
\newcommand{\canOP}[2]{#1 \boxplus #2}
\newcommand{\eOP}[3]{\Bra{ #1,#2,#3}}

\newcommand{\FoxHT}[5]{\mathbbmss{H}^{#1}_{#2}\left\{#4;#3\right\}\left(#5\right)}

\newcommand{\FontDef}[1]{\text{#1}}

\newcommand{\pSeqcdf}{\pSeq_\FontDef{cdf}}

\usepackage{balance}

\usepackage{color}
\usepackage{xcolor}
\usepackage{dsfont}
\usepackage{bbm}
\usepackage[mathscr]{euscript}
\usepackage{braket}
\usepackage{mathtools}

\DeclareMathAlphabet{\mathpzc}{OT1}{pzc}{m}{it}

\DeclareMathAlphabet{\mathitsf}{OML}{cmbr}{m}{it}

\definecolor{CCTLABgreen}{RGB}{0,128,0}

\AtBeginDocument{}

\AtBeginDocument{}

\newtheorem{theorem}{Theorem}

\newtheorem{corollary}{Corollary}
\newtheorem{proposition}{Proposition}
\newtheorem{example}{Example}
\newtheorem{remark}{Remark}

\newtheorem{keynote}{Keynote}

\newtheorem{problem}{Problem}

\DeclareMathOperator{\E}{\mathds{E}}
\DeclareMathOperator{\var}{\mathds{V}\mathrm{ar}}
\DeclareMathOperator{\prob}{\mathds{P}}

\newcommand{\usertext}[1]{\mathrm{#1}}

\newcommand{\R}{\mathbbmss{R}}
\newcommand{\C}{\mathbbmss{C}}
\newcommand{\Z}{\mathbbmss{Z}}
\newcommand{\B}[1]{\mathbf{#1}}

\providecommand\BB{}
\renewcommand{\BB}[1]{\pmb{#1}}

\newcommand{\EX}[1]{\E\left\{{#1}\right\}}

\newcommand{\PDF}[2]{p_{{#1}}\left({#2}\right)}
\newcommand{\CDF}[2]{F_{{#1}}\left({#2}\right)}

\newcommand{\Var}[1]{\var\left\{{#1}\right\}}
\newcommand{\Prob}[1]{\prob\left\{{#1}\right\}}

\newcommand{\Binom}[2]{\usertext{Binom}\left({#1},{#2}\right)}
\newcommand{\Erl}[2]{\usertext{Erl}\left(#1,#2\right)}
\newcommand{\GaV}[2]{\usertext{Gamma}\left(#1,#2\right)}

\newcommand{\PV}[1]{\usertext{Poisson}\left(#1\right)}

\newcommand{\GGaV}[3]{\usertext{GG}\left({#1},{#2},{#3}\right)}

\newcommand{\rayleigh}[1]{\usertext{Rayleigh}\left(#1\right)}
\newcommand{\BP}[3]{\usertext{BP}\left({#1},{#2},{#3}\right)}

\newcommand{\RG}[2]{\mathscr{N}\left({#1},{#2}\right)}

\newcommand{\NB}[2]{\usertext{NB}\left({#1},{#2}\right)}

\newcommand{\argmax}{\mathop{\mathrm{arg\,max}}}

\newcommand{\bTRe}{\begin{dingautolist}{182}}
\newcommand{\eTRe}{\end{dingautolist}}

\newcommand{\cR}{\mathcal{R}}
\newcommand{\GF}[1]{\Gamma\left(#1\right)}

\newcommand{\Tb}{T_\mathrm{b}}
\newcommand{\Pb}[1]{P_{\mathrm{b},{#1}}}
\newcommand{\IPb}[1]{\tilde{P}_{\mathrm{b},{#1}}}

\newcommand{\denI}{\rv{\lambda}}

\newcommand{\pSeqFPTpdf}[1]{\pSeq_{\mathrm{fpt},{#1}}}

\newcommand{\pSeqber}{\pSeq_{\mathrm{ber}}}

\newcommand{\pSeqdetcdf}{\pSeq_{q}}
\newcommand{\pSeqdetcdfb}{\dot{\pSeq}_{q}}
\newcommand{\pSeqImean}{\pSeq_{\mu}}

\newcommand{\aeq}{\mathop{\sim}\limits^{.}}

\begin{document}

\title{
	Molecular Communication with Anomalous Diffusion in Stochastic Nanonetworks
	}
\author{
	\vspace{1cm}
        Dung Phuong Trinh, 
        Youngmin~Jeong, 
        Hyundong~Shin, 
        and
        Moe~Z.~Win 

\thanks{
        D.~P.~Trinh and H.~Shin are with the Department of Electronic Engineering,
        Kyung Hee University,
        1732 Deogyeong-daero, Giheung-gu,
        Yongin-si, Gyeonggi-do 17104 Korea
        (e-mail: \{dungtp, hshin\}@khu.ac.kr).
}
\thanks{Y.~Jeong and M.~Z.~Win are with
            the Laboratory for Information and Decision Systems (LIDS),
            Massachusetts Institute of Technology,
            77 Massachusetts Avenue,
            Cambridge, MA 02139 USA.
            (e-mail: \{ymjeong, moewin\}@mit.edu).
}
}

\maketitle 

\begin{abstract}

Molecular communication in nature can incorporate a large number of nano-things in nanonetworks as well as demonstrate how nano-things communicate. This paper presents molecular communication where transmit nanomachines deliver information molecules to a receive nanomachine over an anomalous  diffusion channel. By considering a random molecule concentration in a space-time fractional diffusion channel, an analytical expression is derived for the first passage time (FPT) of the molecules. Then, the bit error rate of the $\ell$th nearest molecular communication with timing binary modulation is derived in terms of Fox's $H$-function. In the presence of interfering molecules, the mean and variance of the number of the arrived interfering molecules in a given time interval are presented. Using these statistics, a simple mitigation scheme for timing modulation is provided. The results in this paper provide the network performance on the error probability by averaging over a set of random distances between the communicating links as well as a set of random FPTs caused by the anomalous diffusion of molecules. This result will help in designing and developing molecular communication systems for various design purposes.

\end{abstract}


\newpage

\section{Introduction}		\label{sec:1}


\IEEEPARstart{T}{he} internet of things (IoT) is rapidly gaining attention as a new paradigm in the modern field of communications and networks, where the \emph{things}---all types of physical elements, e.g., sensors, tags, electronic devices, mobile phones, and home appliances---are capable of interconnecting with a large number of networks for various applications such as machine communication, smart cities, and intelligent transportation \cite{AIM:10:CN}. As the demands of IoT continue to grow towards a hyper-connected world and the internet of everything, recent developments in nanotechnology have promised that nano solutions compose a new concept of IoT---called the \emph{internet of nano-things} (IoNT)---by using biologically embedded computing devices \cite{DF:15:NCN,APBK:15:COM}. 
However, realizing IoNT requires developing new communication and networking techniques and solving various technical challenges \cite{APBK:15:COM}. 

Molecular communication is a new communication paradigm for transmitting information between machines that are typically a few nanometers to a few micrometers in size, where the information is carried using molecules in a nanonetwork. This new communication system is expected to be practical for  use in various IoNT applications such as drug delivery systems, healthcare systems, nano-materials, and nano-machinery \cite{AFSFH:12:MWCOM,NSOMV:14:NB, AAB:12:COM,CABK:15:BME,BABK:16:NB}. 

Brownian motion (normal diffusion) has been widely used for ideal diffusion environments since the free movement of molecules is well described in a fluid medium \cite{SEA:12:IT,PA:14:COM,LZMY:17:CL,JAJSS:16:COM,GA:16:COM}. However, various potential applications of molecular communication cannot be limited to those ideal environments and we may meet extraordinary diffusion in crowded, heterogenous, and complex structure environments, e.g., water molecules in human tissue, turbulent plasma, bacterial motion, amorphous semiconductors, the porous system, and the polymeric system 
\cite{WEKN:04:B_J, SSS:97:BJ,MK:00:PR,OBSTVB:06:MR, MJD:09:BJ}. %

The extraordinary diffusion phenomenon was first discovered by Lewis F. Richardson in 1926 in his large volume of experimental data, and this so-called \emph{anomalous diffusion} does not obey normal diffusion theory \cite{Ric:26:PRSA}.\footnote{The terminology \emph{``non-Fickian diffusion''}  first referred to a representation of the modified Fick's second law of diffusion equation.} It was shown that the random propagation of molecules no longer depends on time $t$ linearly but instead time $t^{3/2}$ in a turbulent medium. Since the late 1960s, many researchers have been interested in examining this diffusion for various propagation environments 
\cite{PS:77:PRB,GNZ:85:PRL}, and some mathematical models were built in the presence or absence of an external velocity or force field to describe anomalous dynamic behavior (see, 
\cite{MK:00:PR,TSMD:12:PSA, CTJS:15:CL,MMM:16:NB,MEDR:17:CL}, 
and references therein). Subdiffusion is used to explain the divergence property of waiting time with finite moments of the jump length distribution of the particles. It has been found in various contexts---e.g., the movement of lipids in membranes, cytoplasmic macromolecules in living cells, proteins in the nucleoplasm, and the translocation of polymers \cite{SSS:97:BJ,WEKN:04:B_J}---and the mean squared displacement of molecules scales slower than a linear relation in time.
For a finite mean waiting time and divergent jump length variance of particles, superdiffusion (also known as \emph{L\'{e}vy flights}) has been explored in \cite{MK:00:PR}, which can be observed in turbulent flows or bacterial motions \cite{OBSTVB:06:MR,MJD:09:BJ}. The mean square displacement of superdiffusing molecules increases more rapidly in time than for normal diffusion.

In the context of molecular communication, anomalous diffusion can appear when the concentration of molecules is very high since the collisions between molecules lead to anomalous movement of the  molecules in a given medium. For example, calcium signaling based molecular communication \cite{Cal:07:Cell,KTE:12:MCOM} cannot avoid anomalous diffusion since calcium ions interact with each other due to  the electrostatic forces. Furthermore, experimental studies of molecular communication showed that the channel response is nonlinear and does not obey theoretical results from previous works \cite{FKEC:14:JSAC}. These results motivate the use of  extraordinary diffusion in molecular communication for many applications \cite{CTJS:15:CL, MMM:16:NB, MEDR:17:CL}.

Since the molecular system can consist of a vast number of molecules, it is difficult to characterize the dynamic behavior of the system analytically. Specifically, the modeling of a dynamic concentration (density) of molecules that undergo absorption, reaction, elastic collision and libration is challenging when  designing a dynamic nanonetwork for molecular communication.
Over the last decade, extensive work on molecular communication systems has spurred researchers to propose diverse solutions for how to deliver information in diffusive propagation, where transmit nanomachines (TNs) emit information molecules depending on their encoding scheme 
\cite{BBJK:14:NANO,SEA:12:IT,PA:14:COM,SML:15:WCOM,KYTA:12:NCN,NCS:14:JSAC,CTJS:15:CL,MMM:16:NB}. 
\emph{Co-channel interference} introduces inevitable uncertainty into the diffusion-based molecular nanonetwork when multiple TNs emit molecules simultaneously \cite{PA:14:COM,KYTA:12:NCN}. These interfering molecules can lead to dynamic variation of the molecule concentration in the nanonetwork and degrade the performance of molecular communication. For example, the concentration of interfering molecules at reference time $t_1$ is different from that at time $t_0$ ($t_0 < t_1$) depending on the nanonetwork environment \cite{JCM:10:AEM}. The density of TNs also can vary in the medium. The moving TNs governed by the law of diffusion lead to dynamic changes in the number of TNs.
Therefore, it is crucial to model the dynamic concentration of TNs and interfering molecules in a stochastic way
\cite{DNGNE:17:MBSC,PA:11:SP,PA:14:COM,LHNLC:16:NB,LHLCJ:17:MNL}. 
A statistical-physical model of interference in nanonetworks was introduced in \cite{PA:14:COM} where the co-channel TNs are randomly distributed according to a homogeneous Poisson point process (PPP). The expected number of interfering molecules at the receive nanomachine (RN) has been analyzed under a stochastic geometry framework \cite{DNGNE:17:MBSC}. However, to the best of the authors knowledge, there is no literature considering a \emph{general diffusion channel model} for heterogeneous propagation of molecules considering the dynamic behavior of \emph{random locations of molecules} in large-scale nanonetworks.

In wireless networks, the PPP has been shown to be a good model for random positions of communicating nodes 
\cite{ESAW:17:CSTO,RQSW:09:JSAC,NJQTS:13:WCOM,JQKS:14:WCOM,TJS:17:ACCESS}.
This spatial model is fully described by the spatial (deterministic) density. However, this model often fails to capture the network dynamics arising from node mobility, the network geometry, and network scheduling in space and time. In traffic theory, the traffic flow is well fitted to a negative binomial distribution for high-variant traffic in the space and time domains \cite{GH:75:Book}. Specifically, the gamma-distributed TN concentration can explain the cyclic-variants and dense concentration scenarios. Hence, we consider a versatile family of statistical distributions for a general distributional structure of the molecule concentration in a stochastic spatial model.

In this paper, we consider molecular communication in a stochastic nanonetwork. Specifically, we are interested in characterizing the performance of the $\ell$th nearest molecular communication from the viewpoint of the network rather than the performance of specific communication links. To this end, we embody the spatial randomness of TNs and interfering molecules according to a stochastic process with random distances between communicating nanomachines. To account for the extraordinary propagation of molecules, we consider anomalous diffusion based on the space-time fractional diffusion equation, which encompasses various types of diffusion scenarios, including Brownian motion. The main contributions of this paper can be summarized as follows.

\begin{itemize}
\item

We characterize the first passage time (FPT) in the $\left(\alpha,\beta\right)$-anomalous diffusion with random distances determined by the $H$-molecule concentration (see Proposition~\ref{pro:1}). We first derive the probability density function (PDF) of the FPT, where the $H$-variate is chosen for an arbitrarily distributed random distance ($H$-distance) between the information molecule emitted from the stochastic field of TNs in the region $\cR$ and the RN. Due to the Mellin convolution operators of the $H$-function, the FPT with random distances is  again an $H$-variate (see Theorem~\ref{proposition:FPT}). The $\left(\alpha,\beta\right)$-anomalous diffusion encompasses various diffusion scenarios depending on the diffusion parameters $\alpha$ and $\beta$. We particularize the statistical properties of FPT for the normal diffusion with and without spatial randomness of molecule locations (see Remark~\ref{rem:1}).

\item

In the absence of interfering molecules, we analyze the bit error rate (BER) for molecular communication between the $\ell$th nearest TN and the RN with timing modulation when the distance between each TN and the RN is perfectly estimated and known by the RN (see Theorem~\ref{thm:ber:timing}). Then, we provide a BER expression in terms of a single $H$-function when the RN uses a fixed detection threshold (see Theorem~\ref{col:berbound}). This enables us to evaluate the BER while neither estimating the distance nor determining the optimal detection threshold (see Remark~\ref{rem:berbound}). We further show that the low-rate slope of the BER curve is a function of the diffusion parameters $\alpha$ and $\beta$, and a subset of the $H$-parameters of the random distance (see Corollary~\ref{col:slope}). Specifically, the low-rate slope depends only on the diffusion parameters $\alpha$ and $\beta$ in the Poisson field of TNs, and the Cox $\left(a,b\right)$-gamma field of TNs when the shape parameter $a>0.5$ (see Remark~\ref{rem:slope}). 

\item

In the presence of interfering molecules, we characterize the effect of interference on the BER for molecular communication between the $\ell$th nearest TN and the RN in $\left(\alpha,\beta\right)$-anomalous diffusion. Applying Campbell's theorem, we characterize the mean and variance of the number of interfering molecules arriving in a given time interval (see Theorem~\ref{thm:aveint}). 
Since the interfering molecules significantly degrade the BER in timing modulation (see Theorem~\ref{thm:ber:withoutI}), we propose a simple mitigation scheme using the mean and variance of the number of interfering molecules (see Theorem~\ref{thm:ber:timing:withID}). It is shown that BER  degradation depends  on the variance of the number of interfering molecules (see Remark~\ref{rem:var:I}). 

\end{itemize}

The rest of this paper is organized as follows. In Section II, we present the stochastic nanonetwork model and anomalous diffusion channel model based on the space-time fractional diffusion equation. The FPT is analyzed in Section III with a general random distance distribution. In Sections IV and V, we characterize the BER for the $\ell$th nearest molecular communication with and without interfering molecules, respectively. Finally, conclusions are given in Section VI.

\emph{Notation:} Throughout the paper, we shall adopt notation in which random variables are displayed in sans serif, upright fonts; and their realizations in serif, italic fonts. 
We collect the glossary of notation and symbols used in the paper in Appendix. Readers who are not familiar with
the $H$-function, $H$-variate, or $H$-transform can find their basic identities and properties in \cite{JSW:15:IT}.

\begin{figure}[t!]
    \centerline{\includegraphics[width=0.75\textwidth]{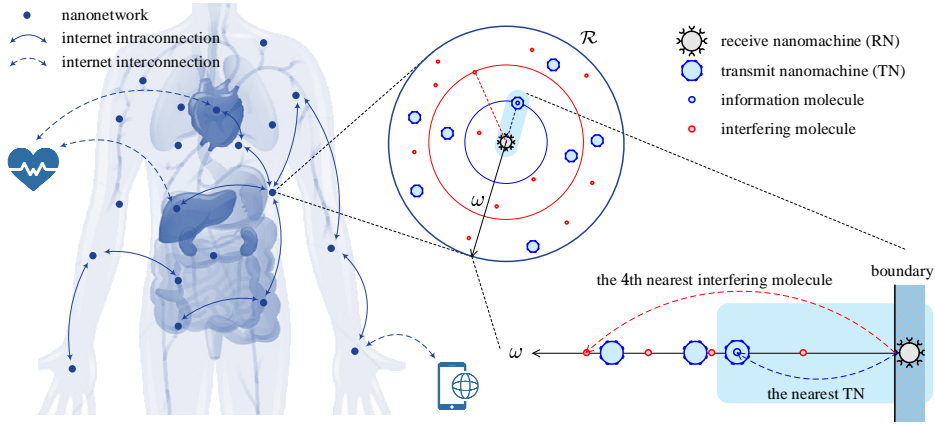}}
    \caption{
        Architecture for the IoNT and an example of a realized nanonetwork in a circular region $\cR$ of radius $\omega$ according to stochastic fields of the TNs and interfering molecules, which obey the anomalous diffusion law. 
    }
    \label{fig:1}
\end{figure}

\section{System Model}

We consider a stochastic nanonetwork, as illustrated in Fig.~\ref{fig:1}, where TNs that are diffused continuously in a two-dimensional region $\cR$ emit molecules to deliver information to an RN located at the origin in the presence of interfering molecules randomly scattered through out the space.\footnote{Since the molecules emitted from the TN  deliver information to the RN via a diffusion process, our framework can also be extended to a stochastic network model under the assumption of spatial random deployment of information molecules.} We assume that the RN acts as a perfect absorbing boundary \cite{SEA:12:IT}; hence, our attention can be focused on the distances between TNs and the RN, considering that the FPT of molecules depends on the random distances between TNs and RN. The random distance between a TN and the RN, which depends on the spatial dimensions and the stochastic process of TNs, can be found using the mapping theorem \cite{SKM:96:Book}.

\subsection{Stochastic Nanonetwork Model}

The TNs and interfering molecules are assumed to be scattered according to stationary Cox processes in the two-dimensional plane $\R^2$. Specifically, we consider that the random intensity $\rv{\lambda}$ of a Cox process is an $H$-variate with $\rv{\lambda}\sim\FoxV{\pM}{\pN}{\pP}{\pQ}{\pSeq=\left(\pK,\pC,\aV,\bV,\sV,\tV\right)}$. We consider a fixed average intensity of TNs or interfering molecules in space. This is valid when the topology of nanonetwork rapidly changes in a short time due to the high mobility of TNs or molecules while keeping their average intensities. This scenario is also valid when molecules in the medium can be degraded by chemical reactions \cite{DNGNE:17:MBSC}.
We begin by introducing a \emph{random molecule concentration} and the corresponding \emph{ordered random distance} to capture the dynamics of the nanonetworks. These are invoked to develop the analysis framework for the $\ell$th nearest molecular communication, e.g., the FPT with the random distance and interference characteristics in nanonetworks.

\begin{proposition}[$H$-Molecule Concentration] \label{pro:1}
Let $\rv{\lambda}\sim \FoxV{\pM}{\pN}{\pP}{\pQ}{\pSeq}$ be the random molecule density. Then the probability of $\ell$ molecules inside the region $\cR$, $\ell\in\Z_+$, is given by
\begin{align} \label{eq:Prolmol}
\Prob{\ell~\text{molecules in}~\cR}
&
    =
    \Fox{\pN+1}{\pM}{\pQ}{\pP+1}
    {\left|\cR\right|;
     \left(
            \frac{\pK}{\left|\cR\right|\ell!},
            \frac{1}{\pC},
            \B{1}_{\pQ}-\bV,
            \left(1+\ell,\B{1}_{\pP}-\aV\right),
            \tV,
            \left(1,\sV\right)
     \right)
    }.
\end{align}
The distance of the $\ell$th nearest molecule from the origin, denoted by $\rv{r}_\ell$, is the $H$-variate $\rv{r}_\ell \sim \FoxV{\pN+1}{\pM}{\pQ}{\pP+1}{\pSeq_\ell}$, where $\pSeq_\ell$ is given by
\begin{align} \label{eq:pSeql}
     \pSeq_\ell
     &=
     \left(
            \frac{\pK\sqrt{\pi}}{\pC^{\frac{3}{2}}\left(\ell-1\right)!},
            \sqrt{\frac{\pi}{\pC}},
            \B{1}_{\pQ}-\bV-\frac{3}{2}\tV,
            \left(\ell-\frac{1}{2},\B{1}_{\pP}-\aV-\frac{3}{2}\sV\right),
            \frac{1}{2}\tV,
            \left(\frac{1}{2},\frac{1}{2}\sV\right)
     \right).
\end{align}
\begin{proof}
The proof is an almost verbatim copy of the proof of \cite[Theorem~1]{JCSW:13:JSAC} in a two-dimensional Cox field of molecules.
\end{proof}
\end{proposition}
We can make the following remarks on the molecule density $\rv{\lambda}$ for special cases.

\begin{itemize}
\item
    \emph{Gamma Molecule Concentration (Cox $\left(a,b\right)$-Gamma):} Let $\rv{\lambda}\sim \GaV{a}{b}$ and $\rv{v}\left(\cR\right)$ be the number of molecules inside the region $\cR$. Then $\rv{v}\left(\cR\right)$ is the negative binomial variable
    \begin{align} 
        \rv{v}\left(\cR\right) \sim \NB{a}{\frac{b\left|\cR\right|}{b\left|\cR\right|+1}}
    \end{align}
    and the distance of the $\ell$th nearest molecule from the origin is the $H$-variate $\rv{r}_\ell \sim \FoxV{1}{1}{1}{1}{\pSeq_\ell}$, where    
    \begin{align}
    \pSeq_\ell
    &=
    \left(
            \frac{\sqrt{\pi b}}{\GF{a}\left(\ell-1\right)!},
            \sqrt{\pi b},
            -a+\frac{1}{2},
            \ell-\frac{1}{2},
            \frac{1}{2},
            \frac{1}{2}
    \right).
    \end{align}

\item   \emph{Deterministic Molecule Concentration (Poisson):}
    When the molecule density has a deterministic concentration, the Cox process boils down to a homogeneous PPP. Let
    $\rv{\lambda}\sim \GaV{a}{b=\lambda_0/a}$. Then as $a\rightarrow \infty$, we have $\rv{\lambda}=\lambda_0$ with probability one. Hence, the number of molecules $\rv{v}\left(\cR\right)$ and the distance of the $\ell$th nearest molecule from the origin are respectively
    \begin{align}
        \rv{v}\left(\cR\right)
        &\sim \PV{\lambda_0 \left|\cR\right|}
    \end{align}
    and
    \begin{align}
    \rv{r}_\ell
    &\sim
        \FoxV{1}{0}{0}{1}
        {\pSeq_\ell=
         \left(
                \frac{\sqrt{\lambda_0\pi}}{\left(\ell-1\right)!},
                \sqrt{\lambda_0\pi},
                -,
                \ell-\frac{1}{2},
                -,
                \frac{1}{2}
         \right)
        }
    \nonumber \\
    &=
    \GGaV{\ell}{1/\sqrt{\lambda_0\pi}}{2}.
    \end{align}
    Note that for $\ell=1$, we have \cite[Remark~(a)]{Hae:05:IT}
    \begin{align}
    \rv{r}_1
    &\sim 
        \FoxV{1}{0}{0}{1}
    {
    \pSeq_1=
         \left(
                \sqrt{\lambda_0\pi},
                \sqrt{\lambda_0\pi},
                -,
                \frac{1}{2},
                -,
                \frac{1}{2}
         \right)
    }
    \nonumber \\
    &=
    \rayleigh{1/\sqrt{2\lambda_0 \pi}}
     \end{align}
    as expected.\footnote{It follows from the mapping theorem \cite{SKM:96:Book} that the squared distance for $\cR$  follows the Erlang distribution $\rv{r}_\ell^2 \sim \Erl{\ell}{\lambda_0\pi}$ and represents Poisson arrivals on the line $\R_+$ with the arrival rate $\lambda_0 \pi$. For the gamma molecule density, it can be interpreted as the \emph{Compound gamma} arrivals on the line $\R_+$ with the scale parameters $a$ and $\pi b$, following the beta prime distribution (beta distribution of the second kind) $\rv{r}_\ell^2\sim \BP{\ell}{a}{\pi b}$.
    }
\end{itemize}

\subsection{Anomalous Diffusion Channel Model}

\begin{figure}[t!]
    \centerline{\includegraphics[width=0.7\textwidth]{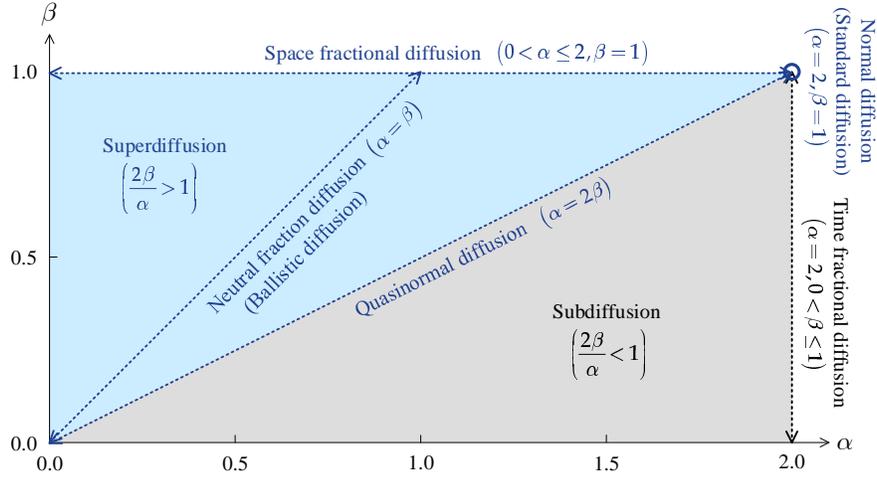}}
    \caption{
    Various types of diffusions in the $\left(\alpha,\beta\right)$-domain.    
    }
    \label{fig:2}
\end{figure}

We consider an $\left(\alpha,\beta\right)$-anomalous diffusion propagation based on a space-time fractional diffusion equation without skewness (or asymmetry) such that
\begin{align}   \label{eq:stfd}
    \frac{\partial^{\beta}}{\partial t^\beta}w\left(x,t\right)
    &=
        K
        \frac{\partial^\alpha}{\partial\left|x\right|^\alpha}
        w\left(x,t\right)
\end{align}
where $w\left(x,t\right)$ is a fundamental solution; $K$ is the diffusion coefficient; $0 < \alpha \leq 2$ is related to the divergence of jump length; and $0 < \beta \leq 1$ is related to the waiting time divergence. With the boundary conditions $w\left(\pm \infty, t\right)=0$ for $t > 0$ and an initial condition $w\left(x,0\right)=\delta\left(x\right)$, the solution of \eqref{eq:stfd} for $\alpha \geq \beta$ is given by \cite{MK:00:PR,MPS:05:CAM,MLP:01:FCAA,TSMD:12:PSA,CTJS:15:CL}\footnote{We obtain the solution \eqref{eq:stfd:sol} by introducing the Caputo derivative \cite{CTJS:15:CL,TSMD:12:PSA}. Note that when $\alpha < \beta$, the $H$-function representation of the fundamental solution can be found in \cite[eq.~(4.2)]{MPS:05:CAM}, which has a different form from \eqref{eq:stfd:sol} due to the singularity of the $H$-function at $x \rightarrow \infty$.} 
\begin{align}   \label{eq:stfd:sol}
    w\left(x,t\right)
    &=
        \frac{1}{\alpha\left|x\right|}
        \FoxH{2}{1}{3}{3}
        {\frac{\left|x\right|}{K^{1/\alpha}t^{\beta/\alpha}}}
        {\left(1,\frac{1}{\alpha}\right),\left(1,\frac{\beta}{\alpha}\right),\left(1,\frac{1}{2}\right)}
        {\left(1,1\right),\left(1,\frac{1}{\alpha}\right),\left(1,\frac{1}{2}\right)}.
\end{align}
The solution $w\left(x,t\right)$ represents a probability density of the molecule location $x$ at a given time $t$. The $\left(\alpha,\beta\right)$-anomalous diffusion can encompass subdiffusion ($\frac{2\beta}{\alpha}<1$), superdiffusion ($\frac{2\beta}{\alpha}>1$), and normal diffusion ($\frac{2\beta}{\alpha}=1$) scenarios depending on the mean squared displacement in the asymptotic limit of large $t$ as $\left<\Delta x^2\right>\propto t^{\frac{2\beta}{\alpha}}$.\footnote{The term ``quasinormal diffusion'' would be more appropriate when $\alpha=2\beta$ with $\alpha < 2$ and $\beta<1$. In this case, the spatial jump length and the waiting time do not lead to Gaussian and Markovian properties, respectively.}  The $\left(\alpha,\beta\right)$-anomalous diffusion can also be classified into particular cases---namely, standard diffusion ($\alpha=2$, $\beta=1$), space fractional diffusion ($0<\alpha \leq 2$, $\beta=1$), time fractional diffusion ($\alpha=2, 0<\beta< 1$), and neutral fractional diffusion ($\alpha=\beta$) \cite{MPS:05:CAM,MLP:01:FCAA}.\footnote{Anomalous diffusion can be characterized by $2\beta/\alpha$, called a \emph{diffusion exponent}, which was measured by $0.84$ or $0.59$ in the cytoplasm of living cell \cite{WEKN:04:B_J}, $0.7$ in the crowded cellular fluids, and $0.65$ or $0.49$ in the cellular membranes \cite{HF:13:RPP}. Specifically, $\left(3/4,1/2\right)$-anomalous diffusion has been observed in a pressure-gradient-driven turbulence model \cite{NCL:04:PP}.} Fig.~\ref{fig:2} shows various types of diffusions, defined through the diffusion parameters $\alpha$ and $\beta$ in the $\left(\alpha,\beta\right)$-domain.

\section{First Passage Time}

In this section, we derive the density of FPT while accounting for the random locations of molecules and anomalous diffusion propagation. Let $\rv{t}\left(r\right)$ be the FPT defined by the time taken for a molecule at $x=0$ to reach distance $x=r$, $r \in \R_{+}$, for the first time:
\begin{align}   \label{eq:FPT}
    \rv{t}\left(r\right)=\inf\left\{ t: x\left(t\right)=r \right\}.
\end{align}
The FPT plays an important role in molecular communication. Specifically, it behaves as noise (uncertainty) in the random propagation time when the information is encoded via molecules according to the release time or the concentration of the molecules. 
For a given (deterministic) distance $r$ and absorbing boundaries at $x=-\infty$ and $x=r$, the FPT for $\alpha \geq \beta$ in $\left(\alpha,\beta \right)$-anomalous diffusion is the $H$-variate \cite[eq.~(4)]{CTJS:15:CL}
\begin{align}	\label{eq:fpt:detr}
\rv{t}\left(r\right)
	\sim
	\FoxV{1}{2}{3}{3}{\pSeq_0
	\Ket{\frac{r^{\alpha/\beta}}{K^{1/\beta}}}}
\end{align}
where the parameter sequence $\pSeq_0$ is given by
\begin{align} \label{eq:pSeq0}
\pSeq_0
&=
    \biggl(\frac{2}{\alpha},
    1,
    \left(-\frac{\alpha}{\beta},-\frac{1}{\beta},-\frac{\alpha}{2\beta}\right),
    \left(-\frac{1}{\beta},-1,-\frac{\alpha}{2\beta}\right),
    \left(\frac{\alpha}{\beta},\frac{1}{\beta},\frac{\alpha}{2\beta}\right),
    \left(\frac{1}{\beta},1,\frac{\alpha}{2\beta}\right)
    \biggr)
\end{align}
and $\pSeq\ket{a}=\left(\pK/a,\pC/a,\aV,\bV,\sV,\tV\right)$ denotes the scaling operation on the parameter sequence $\pSeq$ for $a \in \R_{++}$ \cite[Property~2]{JSW:15:IT}.

%
\begin{theorem}[First Passage Time with $H$-distance]     \label{proposition:FPT}
Let $\rv{r}_\ell\sim \FoxV{\pM}{\pN}{\pP}{\pQ}{\pSeq_\ell}$ be the random distance of the $\ell$th nearest molecule from the RN. Then, the FPT $\rv{t}_\ell$ for $\alpha \geq \beta$ is the $H$-variate
\begin{align}	 \label{eq:FPT:function}
\rv{t}_\ell
\sim \FoxV{\pM+1}{\pN+2}{\pP+3}{\pQ+3}
    {\pSeqFPTpdf{\ell}\Ket{\frac{1}{K^{1/\beta}}}}
\end{align}
where the parameter sequence $\pSeqFPTpdf{\ell}$ is
\begin{align}   \label{eq:FPT:PDF:seq}
    \pSeqFPTpdf{\ell}
    =
        \biggl(
            \frac{2\pK_\ell}{\alpha\pC_\ell^{1-\alpha/\beta}},
            \pC_\ell^{\alpha/\beta},
            \dot{\aV}_\ell,
            \dot{\bV}_\ell,
            \dot{\sV}_\ell,
            \dot{\tV}_\ell
            \biggr)
\end{align}
with
\begin{align}
\begin{cases}
\dot{\aV}_\ell
=
\bigl(
                -\frac{\alpha}{\beta},
                -\frac{1}{\beta},
                \aV_\ell+\bigl(1-\frac{\alpha}{\beta}\bigr)\sV_\ell,
                -\frac{\alpha}{2\beta}
		\bigr)
\\
\dot{\bV}_\ell
=
		\bigl(
                -\frac{1}{\beta},
                \bV_\ell+\bigl(1-\frac{\alpha}{\beta}\bigr)\tV_\ell,
                -1,
                -\frac{\alpha}{2\beta}
		\bigr)
\\
\dot{\sV}_\ell
=
		\bigl(
                \frac{\alpha}{\beta},
                \frac{1}{\beta},
                \frac{\alpha}{\beta}\sV_\ell,
                \frac{\alpha}{2\beta}
		\bigr)
\\
\dot{\tV}_\ell
=
		\bigl(
                \frac{1}{\beta},
                \frac{\alpha}{\beta}\tV_\ell,
                1,
                \frac{\alpha}{2\beta}
		\bigr).
\end{cases}
\end{align}
\begin{proof}
This follows from the definition and elementary identities of the $H$-transform \cite{JSW:15:IT} and  
\begin{align}   \label{eq:FPT:function}
\PDF{\rv{t}_\ell}{t}
&=
\frac{\beta}{\alpha}
    \FoxHT{2,1}{3,3}
    {\eOP{1}{\frac{\beta}{\alpha}}{0}\left(\pSeq_0\Ket{\frac{1}{K^{1/\beta}}}\right)^{-1}}
    {\Fox{\pM}{\pN}{\pP}{\pQ}{r;\Bra{-\frac{\alpha}{\beta}}\pSeq_\ell}}
    {t^{-\frac{\beta}{\alpha}}}
\nonumber \\
&=
    \frac{\beta}{\alpha}
    \Fox{\pN+2}{\pM+1}{\pQ+3}{\pP+3}
    {t^{-\frac{\beta}{\alpha}};\stdOP{\eOP{1}{\frac{\beta}{\alpha}}{0}\left(\pSeq_0\Ket{\frac{1}{K^{1/\beta}}}\right)^{-1}}{\Bra{-\frac{\alpha}{\beta}}{\pSeq_\ell}}}	
\end{align}
where $\eOP{a}{b}{r}\pSeq$ and $\Bra{r}\pSeq$ denote the elementary and conjugate operations on the parameter sequence $\pSeq$ for $r \in \C$ \cite[Property~3 and Remark~2]{JSW:15:IT}, $\pSeq^{-1}$ denotes the inverse operation on the parameter $\pSeq$ \cite[Property~6]{JSW:15:IT}, and $\stdOP{}{}$ denotes the Mellin operation on the two parameter sequences \cite[Proposition~4]{JSW:15:IT}.
\end{proof}
\end{theorem}

\begin{remark}[Normal Diffusion]    \label{rem:1}
For normal diffusion ($\alpha=2$, $\beta=1$), the FPT in Theorem~\ref{proposition:FPT} reduces to 
\begin{align}   \label{eq:FPT:normal}
\rv{t}_\ell
&\sim
    \mathscr{H}^{\pM,\pN+1}_{\pP+1,\pQ+1}
    \biggl(
    \Bigl(
            \pK_\ell \pC_\ell,
            \pC_\ell^2,
            \left(-2,\aV_\ell-\sV_\ell\right),
            \left(\bV_\ell-\tV_\ell,-1\right),
            \left(2,2\sV_\ell\right),
            \left(2\tV_\ell,1\right)
    \Bigr)
    \Ket{\frac{1}{K}}
    \biggr).
\end{align}
Given $\rv{r}_\ell=r$, the location-conditioned FPT for normal diffusion has the PDF
\begin{align}  \label{eq:ftp:normal1}
    \PDF{\rv{t}_\ell}{t}
    &=
    \frac{K}{r^2}
    \FoxH{2}{1}{3}{3}
    {\frac{r^2}{K t}}
    {\left(2,1\right),\left(2,1\right),\left(2,1\right)}
    {\left(3,2\right),\left(2,1\right),\left(2,1\right)}
    \nonumber \\
   &
 =
    \frac{r}{\sqrt{4\pi K t^3}}
    \exp\left(-\frac{r^2}{4 K t}\right).
\end{align}
Note that with spatial conditioning, the diffusion process becomes a \emph{Wiener process} without drift and its  variance is equal to $2K$.\footnote{The distribution \eqref{eq:ftp:normal1} is a L\'{e}vy distribution, or special cases of the inverse Gamma and Pearson-V distributions. Note that the FPT for the Wiener process with drift follows an inverse Gaussian distribution \cite{SEA:12:IT}.}
\end{remark}

\begin{figure}[t!]
    \centerline{\includegraphics[width=0.55\textwidth]{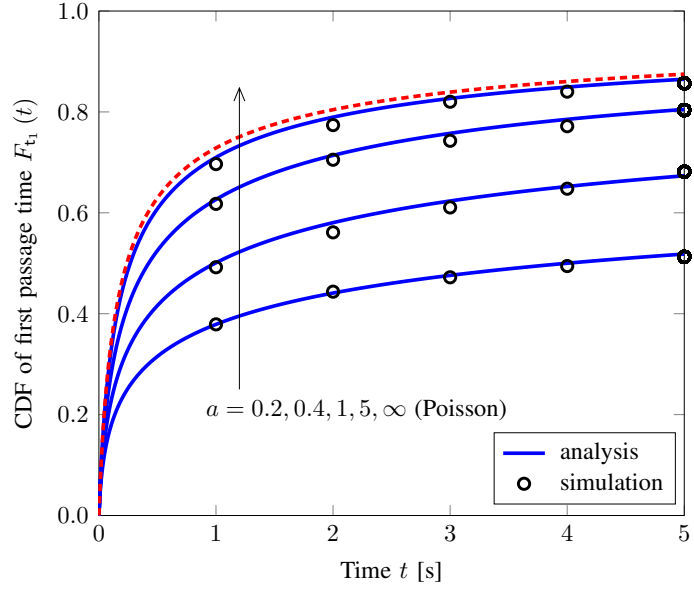}}
\caption{
CDF of the FPT for the nearest molecule in the normal diffusion channel with the Cox $\left(a,10^{10}/a\right)$-gamma field of molecules when $a=0.2,0.4,1,5$ and $\infty$ (Poisson field of molecules).
}
\label{fig:3}
\end{figure}

\begin{figure}[t!]
    \centerline{\includegraphics[width=0.55\textwidth]{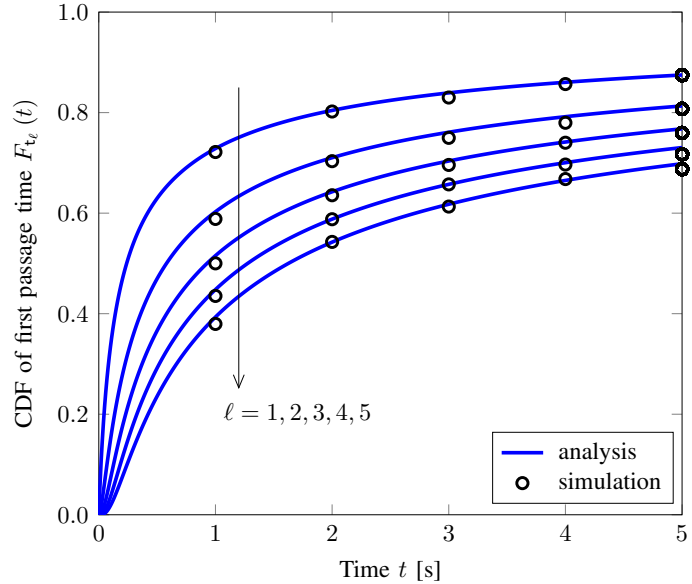}}
\caption{CDF of the FPT for the $\ell$th nearest molecule in the normal diffusion channel with the Poisson field of molecules when $\lambda_0=10^{10}$ [molecules/m$^2$] and $\ell=1,2,3,4,5$.
}
\label{fig:4}
\end{figure}

\begin{example}
To exemplify the FPT between the $\ell$th nearest molecule and the RN in a stochastic field of molecules, we consider two nanonetwork scenarios: i) a Cox $\left(a,b\right)$-gamma field of molecules with a gamma random molecule concentration $\rv{\lambda}\sim \GaV{a}{b}$; and ii) a Poisson field of molecules with a deterministic molecule concentration $\rv{\lambda}=\lambda_0$. We set the diffusion coefficient $K=10^{-10}$~[$\mathrm{m^2/s}$] (for a biological environment) for all examples in this paper. For simulations, we use a Monte Carlo method based on continuous-time random walks \cite{CTJS:15:CL, FSG:08:PRE}. For random FPT in space and in time, 20,000 realizations were used. For each molecule at the initial random position, we used a random discrete time step with a Mittag-Leffler distribution associated with the diffusion parameter $\beta$, and a random new position is obtained by the stable distribution associated with the diffusion parameter $\alpha$ and the diffusion coefficient $K$. Our simulation method can be understood as a generalized version of a particle-based simulator for Brownian motion \cite{AJS:18:COM}.

\begin{figure}[t!]
    \centerline{\includegraphics[width=0.55\textwidth]{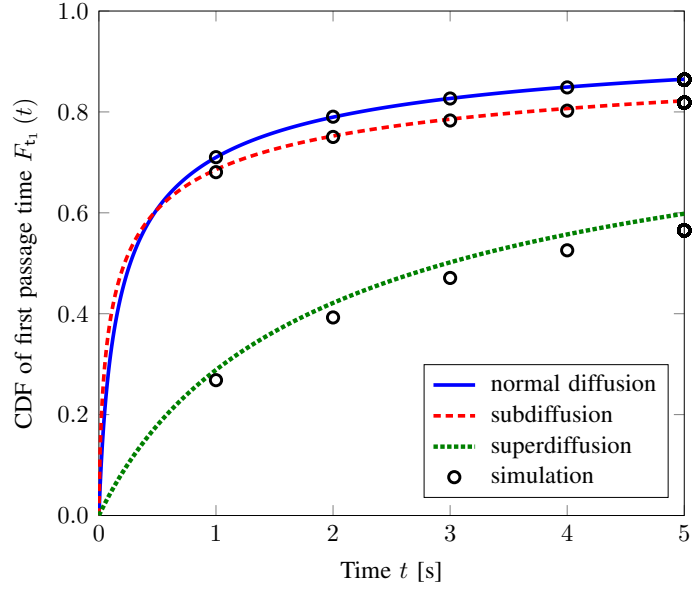}}
\caption{CDF of the FPT for the nearest molecule in the $\left(\alpha,\beta\right)$-anomalous diffusion channel with the Cox $\left(5,0.2\times 10^{10}\right)$-gamma field of molecules when i) $\left(\alpha,\beta\right)=\left(2,1\right)$ for normal diffusion; ii) $\left(\alpha,\beta\right)=\left(2,0.8\right)$ for subdiffusion; and iii) $\left(\alpha,\beta\right)=\left(1.8,1\right)$ for superdiffusion.
}
\label{fig:5}
\end{figure}

Fig.~\ref{fig:3} shows the cumulative distribution function (CDF) of the FPT for the nearest molecule in a  normal diffusion channel with the Cox $\left(a,10^{10}/a\right)$-gamma field of molecules when $a=0.2,0.4,1,5$ and $\infty$ (Poisson field of molecules). In this figure, we set the average molecule concentration as $\EX{\rv{\lambda}}=10^{10}$~[molecules/m$^2$] for comparison. We can observe that the molecules are more dispersed in space with small values of $a$ under the same average molecule concentration. This follows from the fact that the negative binomial arrival (distance) with gamma random concentration exhibits the over-dispersed statistical property. We can also see that the FPT in the Cox gamma field of molecules behaves like that in the Poisson field of molecules with a large value of $a$ due to the loss of randomness in the concentration, as expected. 
To ascertain the spatial ordering characteristic of the FPT in $\cR$, the CDF of the FPT for the $\ell$th nearest molecule in the normal diffusion channel  is depicted in Fig.~\ref{fig:4}, with the Poisson field of molecules with $\lambda_0=10^{10}$ [molecules/m$^2$] and $\ell=1,2,3,4,5$. We can see from the figure that the FPT between the $\ell$th and the $\left(\ell+1\right)$th molecules in the Poisson field exponentially decreases with the spatial ordering index $\ell$.
The extraordinary diffusion effects on the FPT can be ascertained by referring to Fig.~\ref{fig:5}, where the CDF of the FPT for the nearest molecule in the $\left(\alpha,\beta\right)$-anomalous diffusion channel with the Cox $\left(5,0.2\times 10^{10}\right)$-gamma field of molecules is depicted when: i) $\left(\alpha,\beta\right)=\left(2,1\right)$ for normal diffusion; ii) $\left(\alpha,\beta\right)=\left(2,0.8\right)$ for subdiffusion; and iii) $\left(\alpha,\beta\right)=\left(1.8,1\right)$ for superdiffusion. We observe that in general, anomalous diffusion has a large dispersion in propagation compared to normal diffusion. In the superdiffusion scenario, the discordance between the analysis and simulation results comes from the fact that the FPT is overestimated in the simulation due to the long jump lengths of the molecule. This phenomenon can also be interpreted as the first passage leapovers, where the first arrival molecule at the distance $r$ is slower than it is first across $r$ \cite{KLCKM:07:PRL}. 


\end{example}

\section{$\ell$th Nearest Molecular Communication}

In this section, we establish a unifying framework to characterize the effects of spatial randomness of TNs in $\left(\alpha,\beta\right)$-anomalous diffusion for $\ell$th nearest molecular communication without interfering molecules, where the distance between the $\ell$th nearest TN and RN is the $H$-distance $\rv{r}_\ell \sim \FoxV{\pM}{\pN}{\pP}{\pQ}{\pSeq_\ell}$. Specifically, we consider timing binary modulation with a single molecule as an information carrier.

\subsection{Assumptions}

We assume that the release time of molecules is perfectly controlled and synchronized between TNs and the RN \cite{SML:15:WCOM, NCS:14:JSAC, Ata:13:CL, MKHC:02:NAC, SMM:13:CL}.\footnote{The problem of synchronization has been investigated using the behavior of individual cells via inter-cell signaling \cite{MKHC:02:NAC}, and the blind synchronization algorithm has been proposed to estimate the channel delay in diffusion-based molecular communication systems \cite{SMM:13:CL}. Optimal and suboptimal symbol synchronization schemes without a molecular clock are proposed in \cite{JAS:17:NB}. Note that synchronization between TNs is not necessary in this paper since the FPT of the molecules emitted from the $\ell$th nearest TN only depends on the location of the TN \cite{DNGNE:17:MBSC}.} 
Each TN uses different types of molecules for encoding information for orthogonality of the channel uses. The TN also uses different types of molecules for each symbol for the inter-symbol-interference free channel. The RN can distinguish molecules either emitted from each TN or emitted from the same TN in different time intervals, and can wait for a long time until the molecules are absorbed \cite{CTJS:15:CL, SEA:12:IT}. This guarantees preservation of orthogonality among different channel uses and inter-symbol-interference free channels while allowing for a large number of molecule types. That is, the complexity of the system increases as the number of TNs and/or the number of channel uses increases. This assumption can be relaxed by introducing a lifetime of molecules, where the molecules dissipate  immediately after a finite time or with an exponential degradation rate \cite{SML:15:WCOM,GAFYLEC:16:WCOM}. Hence, the same type of molecules can be reused. The lifetime of molecules can be observed when enzymes or other chemical reactions degrade the molecules in the channel \cite{NCS:14:NB1}. Note that various modulation schemes were proposed to reduce the complexity of nanomachines and to mitigate inter-symbol-interference \cite{KIK:15:NB,MGMN:18:COM}. The motion of molecules is independent of the TNs and any boundary. The molecules absorbed by the RN at the FPT no longer affect the nanonetworks \cite{SEA:12:IT,PA:14:COM, SML:15:WCOM, KYTA:12:NCN}. We further assume that the channel state information (such as the diffusion coefficient, anomalous diffusion parameters, and  density of TNs) is perfectly estimated and known by the RN for a reliable detection of the transmitted information \cite{NCS:14:NB,JAS:17:CL,MNES:12:SP,WHL:15:CL, CLY:18:NB}.\footnote{The training-based channel state information estimation method was introduced in \cite{JAJSS:16:COM}. Recently, the non-coherence detection method was proposed in the absence of channel state information at the RN \cite{JFSG:18:COM}. Note that the density and intensity estimation methods in Cox processes were investigated for various applications, e.g., see, \cite{CMT:17:AAS} and the references therein.}
Unless these assumptions are not violated, the considered system is applicable to short-range (from nanometers to millimeters) molecular communication in anomalous diffusion mediums such as cytoplasm of living cell, crowded cellular fluids, and cellular membranes \cite{WEKN:04:B_J, SSS:97:BJ, HF:13:RPP}.  The timing modulation scheme with a single molecule for communication is promising and essential in practice since a complex system is difficult to design and implement via biological nanomachines in nature due to the limited processing capacity of nanomachines \cite{AFSFH:12:MWCOM,NSOMV:14:NB}. In addition, the use of a single molecule can be justified from the assumption of independent motion of molecules emitted from multiple TNs \cite{FMGCEG:19:MBSC}.

\subsection{Bit Error Rate Analysis}


The information is encoded by release time $\rv{s}_\ell \in \left\{0,T_b/2\right\}$ for equally-likely bits `0' and `1', where $T_b$ is the time interval for bit transmission \cite{SEA:12:IT}. Then, the arrival time $\rv{y}_{\mathrm{tm},\ell}$ for the information molecule emitted from the $\ell$th nearest TN can be written as
\begin{align} 	\label{eq:at:time}
    \rv{y}_{\mathrm{tm},\ell} = \rv{s}_\ell + \rv{t}\left(\rv{r}_\ell\right)
\end{align}
where $\rv{r}_\ell$ is the distance between the $\ell$th nearest TN and the RN and $\rv{t}\left(\rv{r}_\ell\right)$ is the FPT of the information molecule emitted from the $\ell$th nearest TN.

Let $\hat{\rv{s}}_\ell$ be the decoded release time of the $\ell$th nearest TN, which can be found using maximum likelihood (ML) detection as follows:
\begin{align} 
  	\hat{\rv{s}}_\ell
	=
	\argmax_{s\in\left\{0,\Tb/2\right\}}  \PDF{\rv{y}_{\mathrm{tm},\ell}|\rv{s}_\ell}{y|s}
\end{align}
where
\begin{align}
    \PDF{\rv{y}_{\mathrm{tm},\ell}|\rv{s}_\ell}{y|s}
    =
	\begin{cases}
    \PDF{\rv{t}\left(\rv{r}_\ell\right)}{y-s}, & \text{if~~}y>s\\
	0, & \text{otherwise}.
	\end{cases}
\end{align}

\subsubsection{With Distance Knowledge at the RN}
We first assume that the distance between each TN and the RN is perfectly estimated and known by the RN. The knowledge of distance between the TN and RN is crucial for optimal functionality of molecular communication. Several distance estimation methods together with signal detection schemes were proposed by measuring the round trip time, signal attenuation \cite{MNES:12:SP}, or concentration-peak time \cite{WHL:15:CL, CLY:18:NB} using feedback signals.

\begin{theorem}[BER for Timing Modulation]  \label{thm:ber:timing}
Let $\rv{r}_\ell\sim \FoxV{\pM}{\pN}{\pP}{\pQ}{\pSeq_\ell}$ and $R=1/\Tb$ [bits/s] be the data rate.  With the knowledge of the random distance $\rv{r}_\ell$ at the RN, the detection threshold, denoted by $\gamma$, is the solution of 
\begin{align}	\label{eq:sol:rth}
    \Fox{1}{2}{3}{3}{\gamma-\frac{1}{2R};\pSeq_0\Ket{\frac{\rv{r}_\ell^{\alpha/\beta}}{K^{1/\beta}}}} 
	=
    \Fox{1}{2}{3}{3}{\gamma;\pSeq_0\Ket{\frac{\rv{r}_\ell^{\alpha/\beta}}{K^{1/\beta}}}}.
\end{align}
Then, the BER $\Pb{\ell}$ of molecular communication between the $\ell$th nearest TN and RN for timing  modulation in $\left(\alpha,\beta\right)$-anomalous diffusion is given in terms of the $H$-transform as follows:
\begin{align}   \label{eq:ber:exact}
    \Pb{\ell}
    &=
     \frac{1}{2}
     \biggl(
                1+
                \frac{\beta}{\alpha}\FoxHT{2,2}{4,4}{\eOP{1}{\frac{\beta}{\alpha}}{0}\pSeqber}
                { \Fox{\pM}{\pN}{\pP}{\pQ}{r;\pSeq_0} }{\gamma^{-\beta/\alpha}}
    \nonumber \\
    &\hspace{1.5cm}
    -
                \frac{\beta}{\alpha}\FoxHT{2,2}{4,4}{\eOP{1}{\frac{\beta}{\alpha}}{0}\pSeqber}
                { \Fox{\pM}{\pN}{\pP}{\pQ}{r;\pSeq_0} }{\left(\gamma-\frac{1}{2R}\right)^{-\beta/\alpha}}
      \biggr)
\end{align}
where the parameter sequence $\pSeqber$ is given by
\begin{align} \label{eq:pSeqBerl}
\pSeqber
&=
    \bigg(
    \frac{2}{\beta},\frac{1}{K^{1/\beta}},
    \B{1}_4,
    \left(\B{1}_3,0\right),
    \left(\frac{\alpha}{\beta},\frac{1}{\beta},1,\frac{\alpha}{2\beta}\right),   
    \left(\frac{\alpha}{\beta},\frac{1}{\beta},\frac{\alpha}{2\beta},\frac{\alpha}{\beta}\right)
    \bigg).
\end{align}
\begin{proof}
For equiprobable bits `0' and `1', the conditional BER $\Pb{\ell}\left(\rv{r}_\ell\right)$ is given by
\begin{align}	\label{eq:cdf:ber:1}
    \Pb{\ell}\left(\rv{r}_\ell\right)
    &=
        \frac{1}{2}\int_{\gamma}^\infty \PDF{\rv{y}_{\mathrm{tm},\ell}|\rv{s}_\ell}{y|0} dy
        + \frac{1}{2}\int_{\frac{\Tb}{2}}^{\gamma} \PDF{\rv{y}_{\mathrm{tm},\ell}|\rv{s}_\ell}{y\left|\frac{\Tb}{2}\right.} dy
    \nonumber \\
    &=
        \frac{1}{2}
        \left(
                1
                +\CDF{\rv{t}\left(\rv{r}_\ell\right)}{\gamma-\frac{\Tb}{2}}
                -\CDF{\rv{t}\left(\rv{r}_\ell\right)}{\gamma}
        \right).
\end{align}
The CDF $\CDF{\rv{t}\left(\rv{r}_\ell\right)}{t}$ is given in \cite[eq.~(7)]{CTJS:15:CL} in terms of the $H$-function as 
\begin{align}	\label{eq:cdf:ber:x}
	\CDF{\rv{t}\left(\rv{r}_\ell\right)}{t}
	&=
		1-
		 \Fox{2}{2}{4}{4}{\rv{r}_\ell^{\alpha/\beta}t^{-1};\pSeqber}.
\end{align}
From \eqref{eq:cdf:ber:1} and \eqref{eq:cdf:ber:x}, we obtain \eqref{eq:ber:exact} in terms of the $H$-transform.
\end{proof}
\end{theorem}

Theorem~\ref{thm:ber:timing} provides the network performance of the BER by averaging over the random distances between the TN and the RN as well as the random FPTs caused by anomalous diffusion of molecules. The distance $\rv{r}_\ell$ should be estimated to find the optimal threshold $\gamma$. 

\subsubsection{Without Distance Knowledge at the RN} Although there exist distance estimation schemes for molecular communication,  it is difficult to estimate the exact distances between moving TNs  and the RN in practice. Instead of measuring the distance as well as calculating the optimal threshold $\gamma$, a fixed detection threshold can be used for a simple but effective scheme, as presented  in the following theorem. In this case, the RN does not require exact distance estimation but instead only needs synchronization between each TN and the RN.

\begin{theorem}[Fixed Detection Threshold]   \label{col:berbound}
Let $\rv{r}_\ell\sim \FoxV{\pM}{\pN}{\pP}{\pQ}{\pSeq_\ell}$. Then, the BER $\hat{P}_{\mathrm{b},\ell}$ of molecular communication between the $\ell$th nearest TN and the RN for timing modulation in $\left(\alpha,\beta\right)$-anomalous diffusion  with the fixed detection threshold $\gamma=\Tb/2$ is given by
\begin{align}   \label{eq:berbound}
    \hat{P}_{\mathrm{b},\ell}
    &=
        \frac{1}{2}
        \Fox{\pN+2}{\pM+2}{\pQ+4}{\pP+4}{2R;\hat{\pSeq}_{\mathrm{ber},\ell}}
\end{align}
where the parameter sequence $\hat{\pSeq}_{\mathrm{ber},\ell}$ is given by
\begin{align}
\hat{\pSeq}_{\mathrm{ber},\ell}
&=
        \biggl(
            \frac{2\pK_\ell}{\alpha\pC_\ell},
            \frac{1}{\pC_\ell^{\alpha/\beta}K^{1/\beta}},
            \ddot{\aV}_\ell,
            \ddot{\bV}_\ell,
            \ddot{\sV}_\ell,
            \ddot{\tV}_\ell
        \biggr)
\end{align}
with
\begin{align}
\begin{cases}
\ddot{\aV}_\ell
=
		\Bigl(
                \B{1}_2,\B{1}_{\pQ}-\bV_\ell-\tV_\ell,\B{1}_2
		\Bigr)
\\
\ddot{\bV}_\ell
=
		\Bigl(
                \B{1}_2,\B{1}_{\pP}-\aV_\ell-\sV_\ell,1,0
		\Bigr)
\\
\ddot{\sV}_\ell
=		
		\Bigl(
                1,\frac{1}{\beta},\frac{\alpha}{\beta}\tV_\ell,1,\frac{\alpha}{2\beta}
		\Bigr)
\\
\ddot{\tV}_\ell
=	
		\Bigl(
                \frac{\alpha}{\beta},
                \frac{1}{\beta},
                \frac{\alpha}{\beta}\sV_\ell,
                \frac{\alpha}{2\beta},
                1
		\Bigr).
\end{cases}
\end{align}
\begin{proof}
For $\rv{r}_\ell\sim \FoxV{\pM}{\pN}{\pP}{\pQ}{\pSeq_\ell}$ with $\gamma=\Tb/2$, we have
\begin{align}	\label{eq:cdf:ber:3}
	\hat{P}_{\mathrm{b},\ell}
	&=
	\frac{1}{2}
	\EX{
	                1
                -\CDF{\rv{t}\left(\rv{r}_\ell\right)}{\frac{\Tb}{2}}
	}
	\nonumber \\
	&=
	\frac{1}{2}
	\left(
	1-
	\int_0^{\Tb/2}
	\int_0^{\infty}
	\PDF{\rv{t}\left(\rv{r}_\ell\right)}{t}
	\PDF{\rv{r}_{\ell}}{r}
	dr dt
	\right)
	\nonumber \\
	&=
	\frac{1}{2}
	\left(1-\CDF{\rv{t}_\ell}{\frac{\Tb}{2}}\right).
\end{align}
The CDF $\CDF{\rv{t}_\ell}{t}$ can be found using \cite[eq.~(87)]{JSW:15:IT} in terms of the $H$-function as 
\begin{align}	\label{eq:cdf:ber:2}
	\CDF{\rv{t}_\ell}{t}
	=
		1-
		 \Fox{\pM+2}{\pN+2}{\pP+4}{\pQ+4}{t;\canOP{\pSeqcdf^{-1}}{\Bra{1}\left(\pSeqFPTpdf{\ell}}\Ket{\frac{1}{K^{1/\beta}}}\right)}
\end{align}
where $\canOP{}{}$ denotes the convolution operation on the two parameter sequences \cite[Proposition~5]{JSW:15:IT} and $\pSeqcdf$ is given in \cite[eq.~(58)]{JSW:15:IT}. From \eqref{eq:cdf:ber:3} and \eqref{eq:cdf:ber:2}, we obtain \eqref{eq:berbound}, which completes the proof.
\end{proof}
\end{theorem}

\begin{remark}  \label{rem:berbound}
Since the optimal threshold $\gamma$ is located near $\Tb/2$ and $\gamma \geq \Tb/2$,
\begin{align}
\Delta = \int_{\frac{\Tb}{2}}^{\gamma} \PDF{\rv{y}_{\mathrm{tm},\ell}|\rv{s}_\ell}{y|0} dy - \int_{\frac{\Tb}{2}}^{\gamma} \PDF{\rv{y}_{\mathrm{tm},\ell}|\rv{s}_\ell}{y\left|\frac{\Tb}{2}\right.} dy
\end{align}
is guaranteed to be positive and can be negligible. Hence, the use of the fixed detection threshold $\gamma=\Tb/2$ results in an upper bound of the achievable BER with timing modulation.  Theorem~\ref{col:berbound} gives three main advantages: i) the RN does not require exact distance information between each TN and the RN; ii) it provides a closed-form expression of the BER $\hat{P}_{\mathrm{b},\ell}$ for the timing modulation in terms of the $H$-function without calculating the optimal detection threshold $\gamma$; and iii) hence, we can characterize the low-rate slope as in Corollary~\ref{col:slope}. The optimal threshold $\gamma$ approaches $\Tb/2$ when i) the PDF $\PDF{\rv{t}\left(\rv{r}_\ell\right)}{t}$ of the FPT is less dispersed, ii) the distance between the TN and RN is small, and iii)  low-rate communication. 
\end{remark}

\begin{corollary}[Low-Rate Slope]   \label{col:slope}
Let 
\begin{align}
\zeta_\ell
    \triangleq
    \lim_{R\rightarrow 0}
    \frac{\log \hat{P}_{\mathrm{b},\ell}}{\log R}
\end{align}
be the \emph{low-rate slope} of the BER $\hat{P}_{\mathrm{b},\ell}$. 
Then, for $\rv{r}_\ell \sim \FoxV{\pM}{\pN}{\pP}{\pQ}{\pSeq}$, we have
\begin{align}
    \zeta_\ell
    =
        \min_{j \in \left\{1,2,\ldots,\pN\right\}}
        \left\{
                \frac{\beta}{\alpha},
                \beta,
                \frac{\beta}{\alpha}
                \left(\frac{1-\pA{j}}{\pS{j}}-1\right)
        \right\}
\end{align}
where $\alpha \geq \beta$.
\begin{proof}
Using the asymptotic expansion of the $H$-function \cite[Proposition~3]{JSW:15:IT} for $\alpha \geq \beta$, we have
\begin{align}
   \hat{P}_{\mathrm{b},\ell}
    &\doteq
    \Fox{\pN+2}{\pM+2}{\pQ+4}{\pP+4}{R;\hat{\pSeq}_{\mathrm{ber},\ell}}
    \nonumber \\
    &\doteq
    \left(1/R\right)^{-
        \min_{j \in \left\{1,2,\ldots,\pN\right\}}
        \left\{
                \frac{\beta}{\alpha},
                \beta,
                \frac{\beta}{\alpha}\left(\left(1-\pA{j}\right)/\pS{j}-1\right)
        \right\}
    }
\end{align}
which completes the proof.
\end{proof}
\end{corollary}

\begin{remark}[Low-Rate Slope]     \label{rem:slope}
The low-rate slope $\zeta_\ell$ is a function of the anomalous diffusion parameters $\alpha$ and $\beta$, and the $\pA{}$-, $\pS{}$-parameters of the $H$-distance.\footnote{Note that $\zeta_\ell>0$ always holds since $\pA{j}+\pS{j}<1$ for $j \in \left\{1,2,\ldots,\pN\right\}$, which are necessary conditions such that the $H$-function is a density function \cite[Remark~7]{JSW:15:IT}.}
For the Cox $\left(a,b\right)$-gamma field of molecules, the low-rate slope is equal to $\zeta_\ell=\min\bigl\{\frac{\beta}{\alpha},\beta,\frac{2\beta}{\alpha}\,a\bigr\}$. Hence when $a > 0.5$ (for the Poisson field of molecules), the low-rate slope only depends on the anomalous diffusion parameters $\alpha$ and $\beta$, not on the spatial ordering index $\ell$ and/or spatial concentration $\rv{\lambda}$. For example, the low-rate slope in normal diffusion for the Cox $\left(a,b\right)$-gamma field of TNs is equal to  $\zeta_\ell=0.5$ for $a > 0.5$ while $\zeta_\ell=a$ for $a < 0.5$.
\end{remark}

\begin{figure}[t!]
    \centerline{\includegraphics[width=0.55\textwidth]{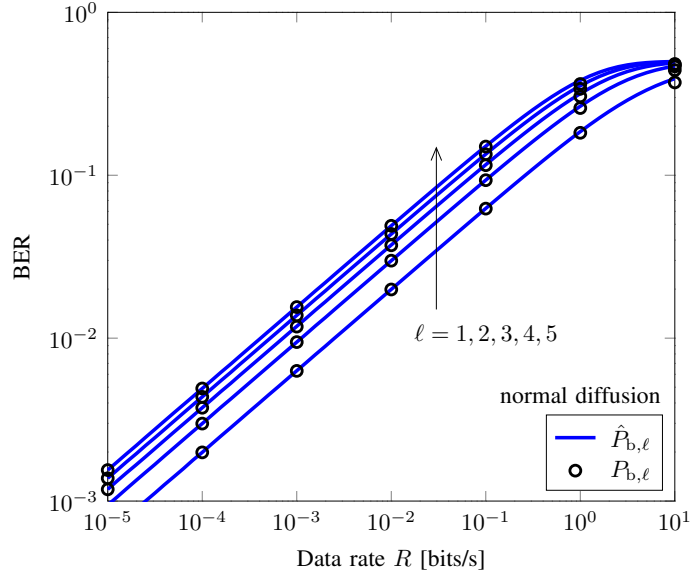}}
\caption{BERs $\Pb{\ell}$ and $\hat{P}_{\mathrm{b},\ell}$ as functions of the data rate $R$ for the $\ell$th molecular communication with timing modulation in the normal diffusion channel with the Poisson field of TNs when $\lambda_0=10^{10}$ [$\mathrm{TNs/m^2}$]  and $\ell=1,2,3,4,5$.
}
\label{fig:6}
\end{figure}

\begin{figure}[t!]
    \centerline{\includegraphics[width=0.55\textwidth]{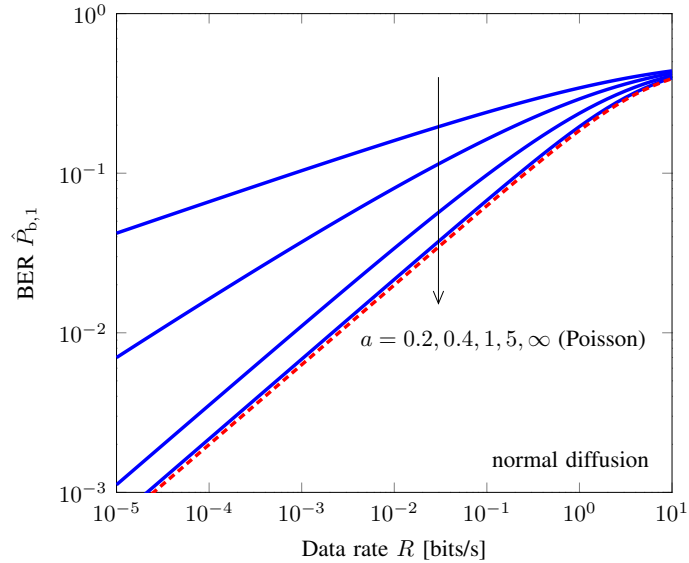}}
\caption{BER $\hat{P}_{\mathrm{b},1}$ as a function of the data rate $R$ for the nearest molecular communication with timing modulation in the normal diffusion channel with the Cox $\left(a,10^{10}/a\right)$-gamma field of TNs when $a=0.2,0.4,1,5$ and $\infty$ (Poisson field of the TNs).
}
\label{fig:7}
\end{figure}

\subsection{Numerical Examples}
\begin{figure}[t!]
    \centerline{\includegraphics[width=0.55\textwidth]{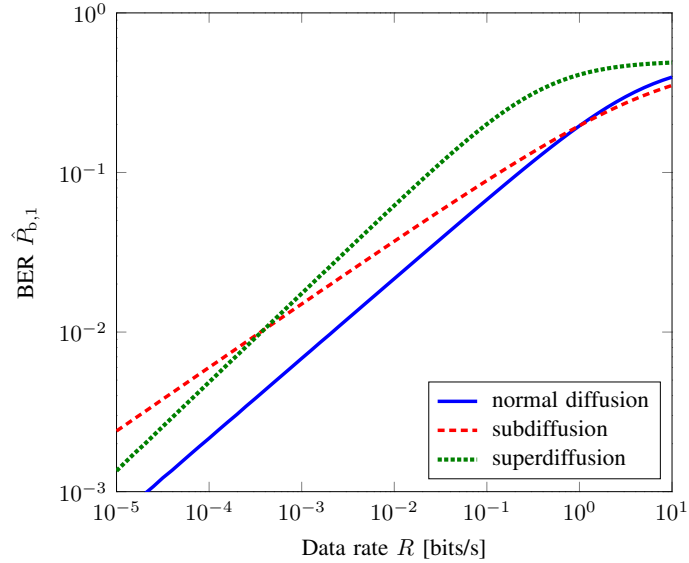}}
\caption{BER $\hat{P}_{\mathrm{b},1}$ as a function of the data rate $R$ for the nearest molecular communication with timing modulation in the $\left(\alpha,\beta\right)$-anomalous diffusion channel with the Cox $\left(5,0.2\times 10^{10}\right)$-gamma field of TNs when: i) $\left(\alpha,\beta\right)=\left(2,1\right)$ for normal diffusion; ii) $\left(\alpha,\beta\right)=\left(2,0.8\right)$ for subdiffusion; and iii) $\left(\alpha,\beta\right)=\left(1.8,1\right)$ for superdiffusion.
}
\label{fig:8}
\end{figure}

Fig.~\ref{fig:6} shows the BERs $\Pb{\ell}$ and $\hat{P}_{\mathrm{b},\ell}$ as functions of the data rate $R$ for the $\ell$th molecular communication with timing modulation in the normal diffusion channel with the Poisson field of TNs when $\lambda_0=10^{10}$ [$\mathrm{TNs/m^2}$] and $\ell=1,2,3,4,5$. 
We can see that the BER $\hat{P}_{\mathrm{b},\ell}$ is extremely tight relative to the BER $\Pb{\ell}$, due to how the detection threshold $\gamma$ approaches $\Tb/2$ in the low-rate regime and/or the TN approaches the RN. We also see that the spatial ordering index $\ell$ does not affect the low-rate slope (Remark~\ref{rem:slope}).
Fig.~\ref{fig:7} shows the BER $\hat{P}_{\mathrm{b},1}$ as a function of the data rate $R$ for the nearest molecular communication with timing modulation in the normal diffusion channel with the Cox $\left(a,10^{10}/a\right)$-gamma field of TNs when $a=0.2,0.4,1,5$ and $\infty$. We observe that the BER decreases with $a$ under the same average concentration due to the smaller amount of dispersion. In this example, the low-rate slopes are equal to $\zeta_1=0.2$ and $0.4$ for $a=0.2$ and $0.4$, respectively, and $\zeta_1=0.5$ for the other cases $a>0.5$, as noted in Corollary~\ref{col:slope} and Remark~\ref{rem:slope}.
To ascertain the effects of anomalous diffusion on the BER, we plot the BER $\hat{P}_{\mathrm{b},1}$ in Fig.~\ref{fig:8} as a function of the data rate $R$ for the nearest molecular communication with timing  modulation in the $\left(\alpha,\beta\right)$-anomalous diffusion channel with the Cox $\left(5,0.2\times 10^{10}\right)$-gamma field of TNs when: i) $\left(\alpha,\beta\right)=\left(2,1\right)$ for normal diffusion; ii) $\left(\alpha,\beta\right)=\left(2,0.8\right)$ for subdiffusion; and iii) $\left(\alpha,\beta\right)=\left(1.8,1\right)$ for superdiffusion. Together with Fig.~\ref{fig:5}, the BER performance of subdiffusion outperforms normal diffusion at a high rate since subdiffusion is less dispersed than normal diffusion. In this example, $\zeta_1=0.50$ for normal diffusion; $\zeta_1=0.40$ for subdiffusion; and $\zeta_1=0.56$ for superdiffusion.

\vspace{-0.1cm}

\section{Communication in the Presence of Interfering Molecules}

In this section, we consider $\ell$th nearest molecular communication in the presence of interfering molecules. 
The interfering molecules may originate from natural sources in the environment. There exist TNs that communicate with another RN using molecules that are of the same type as that used in the $\ell$th nearest molecular communication. Since individual dynamic behaviors of interfering molecules such as generation and extinction are unknown at the RN, we consider interfering molecules scattered in the region $\cR$ according to a Cox process $\rv{\Psi}$ independent of the stochastic field of TNs, denoted by $\rv{\Psi}\left(\cR\right)$ or simply $\rv{\Psi}_\cR$.

\subsection{Interference Characterization}

Let $\rv{z}_T$ be the number of interfering molecules arriving at the RN during the interval $T$. Then, $\rv{z}_T|\rv{\Psi}\left(\cR\right)$ is a \emph{Poisson binomial} variable with respective mean and variance:
\begin{align}
    \mu_T\left(\rv{\Psi}_\cR\right)
    &=
        \sum_{\RV{x} \in \rv{\Psi}\left(\cR\right)}
        q_T\left(\RV{x}\right)
    \\
    \sigma^2_T\left(\rv{\Psi}_\cR\right)
    &=
        \sum_{\RV{x} \in \rv{\Psi}\left(\cR\right)}
        \left(1-q_T\left(\RV{x}\right)\right)
        q_T\left(\RV{x}\right)
\end{align}
where $q_T\left(\RV{x}\right)$ is the probability that the interfering molecule located at $\RV{x} \in \rv{\Psi}\left(\cR\right)$ arrives at the RN during  $T$. For a given $\RV{x}$, the probability $q_T\left(\RV{x}\right)$ can be obtained from the CDF of the FPT in \eqref{eq:fpt:detr} using \cite[eqs.~(86) and (87)]{JSW:15:IT} as follows:
\begin{align}	\label{eq:qti}
    q_T\left(\RV{x}\right)
    &=
    1-
        \Fox{2}{2}{4}{4}{\left\|\RV{x}\right\|^{\alpha/\beta} T^{-1};
        \pSeqdetcdf
        }
   \nonumber \\
   &=
   \Fox{2}{1}{3}{3}{\left\|\RV{x}\right\|^{\alpha/\beta} T^{-1};
        \pSeqdetcdfb
}
\end{align}
where
\begin{align}
        \pSeqdetcdf
        &=
        \bigg(\frac{2}{\beta},\frac{1}{K^{1/\beta}},
    \B{1}_4,
    \left(\B{1}_3,0\right),
    \left(\frac{\alpha}{\beta},\frac{1}{\beta},1,\frac{\alpha}{2\beta}\right),   
    \left(\frac{\alpha}{\beta},\frac{1}{\beta},\frac{\alpha}{2\beta},\frac{\alpha}{\beta}\right)\bigg)
\\
    \pSeqdetcdfb
        &=
        \left(\frac{2}{\beta},\frac{1}{K^{1/\beta}},
    \B{1}_3,
    \left(0,\B{1}_2\right),
    \left(\frac{1}{\beta},1,\frac{\alpha}{2\beta}\right),    
    \left(\frac{\alpha}{\beta},\frac{1}{\beta},\frac{\alpha}{2\beta}\right)\right).
\end{align}

\begin{theorem} \label{thm:aveint}
Let $\omega$ be the radius of $\cR$ and $\rv{\lambda}$ be the random intensity for the Cox process $\rv{\Psi}$ of interfering molecules. Then, the mean and variance of $\rv{z}_T$ are given by
\begin{align}
\EX{\rv{z}_T}
&=
\Var{\rv{z}_T}
=
        4\pi\EX{\rv{\lambda}}
        \frac{T^{\frac{2\beta}{\alpha}}}{\alpha}
        \Fox{2}{2}{4}{4}{\omega T^{-\beta/\alpha};\pSeqImean}
        \label{eq:am:I:mean}
\end{align}
where
\begin{align}	\label{eq:pSeqImean}
\pSeqImean
&=
    \biggl(
            K^{2/\alpha},
            \frac{1}{K^{1/\alpha}},
            \acute{\aV},
            \acute{\bV},
            \acute{\sV},
            \acute{\tV}
    \biggr)
\end{align}
with
\begin{align}
\begin{cases}
\acute{\aV}
=
            \bigl(1,1+\frac{2}{\alpha},1+\frac{2\beta}{\alpha},2\bigr)\\
\acute{\bV}
=         
            \bigl(2,1+\frac{2}{\alpha},2,0\bigr) \\
\acute{\sV}
=            
            \bigl(1,\frac{1}{\alpha},\frac{\beta}{\alpha},\frac{1}{2}\bigr)\\
\acute{\tV}
=            
            \bigl(1,\frac{1}{\alpha},\frac{1}{2},1\bigr).
\end{cases}
\end{align}
\begin{proof}
Using the law of total expectation, we have
\begin{align} 
    \EX{\rv{z}_T}
    &=
    \mathbb{E}\Bigl\{\EX{\rv{z}_T|\rv{\Psi}\left(\cR\right)}\Bigr\}
    =
    \mathbb{E}
    \Biggl\{
        \sum_{\RV{x} \in \rv{\Psi}\left(\cR\right)}
        q_T\left(\RV{x}\right)
    \Biggr\}
    \nonumber \\
    &~
    \mathop = \limits^{\mathrm{(a)}}
        2\pi\EX{\rv{\lambda}}
        \int_0^\omega
        \Fox{2}{1}{3}{3}{r^{\alpha/\beta} T^{-1};\pSeqdetcdfb}
        r dr
    \nonumber \\
    &~
    \mathop = \limits^{\mathrm{(b)}}
        2\pi\EX{\rv{\lambda}}
        \int_0^\omega
        \frac{\beta}{\alpha}
        T^{\beta/\alpha}
        \Fox{2}{1}{3}{3}{r T^{-\beta/\alpha};\eOP{1}{\frac{\beta}{\alpha}}{1}\pSeqdetcdfb}
        dr  
    \nonumber \\
    &~
    \mathop = \limits^{\mathrm{(c)}}
        2\pi\EX{\rv{\lambda}}
        \frac{\beta T^{\frac{2\beta}{\alpha}}}{\alpha}
        \Fox{2}{2}{4}{4}{\omega T^{-\beta/\alpha};\canOP{\pSeqcdf}{\eOP{1}{\frac{\beta}{\alpha}}{2}\pSeqdetcdfb}}
\end{align}
where $\mathrm{(a)}$ follows from Campbell's theorem \cite{SKM:96:Book,HG:08:FTN}; 
 $\mathrm{(b)}$ is obtained from the elementary operation of the $H$-function \cite[Property~5]{JSW:15:IT} and the fact that
\begin{align}
\eOP{1}{\beta_2}{\gamma_2}
\eOP{\alpha}{\beta_1}{\gamma_1}\pSeq
=
	\eOP{\alpha}{\beta_1\beta_2}{\frac{\gamma_1}{\beta_2}+\gamma_2}\pSeq
\end{align}
and $\mathrm{(c)}$ follows from the CDF expression for a $H$-variate \cite[eq.~(86)]{JSW:15:IT}. 
Note that 
\begin{align}
\EX{\sigma_T^2\left(\rv{\Psi}_\cR\right)}
&=
    2\pi\EX{\rv{\lambda}}
    \int_0^\omega
    \Fox{2}{2}{4}{4}{r^{\alpha/\beta} T^{-1};\pSeqdetcdf}
    \Fox{2}{1}{3}{3}{r^{\alpha/\beta} T^{-1};\pSeqdetcdfb}
    r
    dr
    \label{eq:am:I:var}
\\
\Var{\mu_T\left(\rv{\Psi}_\cR\right)} 
&=
        2\pi\EX{\rv{\lambda}}
        \int_0^\omega
        \left[1-\Fox{2}{2}{4}{4}{r^{\alpha/\beta} T^{-1};\pSeqdetcdf}\right]
        \Fox{2}{1}{3}{3}{r^{\alpha/\beta} T^{-1};\pSeqdetcdfb}
        r
        dr     
        \label{eq:var:I:mean} 
\end{align}
again from Campbell's theorem. Hence, using the law of total variance, $\Var{\rv{z}_T}$ is given by
\begin{align} 
	\Var{\rv{z}_T}
	&=
		\EX{\Var{\rv{z}_T|\rv{\Psi}\left(\cR\right)}}
		+
		\Var{\EX{\rv{z}_T|\rv{\Psi}\left(\cR\right)}}
	\nonumber \\
	&=
		\EX{\sigma_T^2\left(\rv{\Psi}_\cR\right)}
		+
		\Var{\mu_T\left(\rv{\Psi}_\cR\right)}
	\nonumber \\
	&=
		\EX{\rv{z}_T}
\end{align}
which completes the proof.
\end{proof}
\end{theorem}

\begin{remark}  \label{rem:inf:time}
As $T \rightarrow \infty$, $\EX{\rv{z}_T}$ and $\Var{\rv{z}_T}$ converge to $\pi \omega^2 \EX{\rv{\lambda}}$ obviously. 
As $\omega \rightarrow \infty$, $\EX{\rv{z}_T}$ and $\Var{\rv{z}_T}$ converge to
\begin{align}   \label{eq:limit:mean:omega}
    2\pi\EX{\rv{\lambda}} \frac{K T^{\beta}}{\GF{1+\beta}}
\end{align}
for $\left(2,\beta\right)$-anomalous diffusion (time-fractional diffusion); increase linearly with $\omega$ for $\left(1,1\right)$-anomalous diffusion; and increase nonlinearly with $\omega$ for other diffusions. 
Note that for both large $\omega$ and $T$, the mean and variance of $\rv{z}_T$ asymptotically scale with $\frac{2\beta}{\alpha}$, i.e., $\EX{\rv{z}_T} \aeq T^{\frac{2\beta}{\alpha}}$ and $\Var{\rv{z}_T} \aeq T^{\frac{2\beta}{\alpha}}$.

\end{remark}

\begin{remark}[Interference in Normal Diffusion]
For normal diffusion, we have 
\begin{align}   \label{eq:am:I:mean:normal}
\EX{\rv{z}_T}
&=\Var{\rv{z}_T} 
\nonumber \\
&
=
        2\pi T\EX{\rv{\lambda}}
        H^{1,1}_{2,2}\Bigl(\omega T^{-1/2};
        \Bigl(K,\frac{1}{K^{1/2}},\left(1,2\right),\left(2,0\right),
        \Bigl(1,\frac{1}{2}\Bigr),\B{1}_2\Bigr).
\end{align}
As $ \omega \rightarrow \infty$, \eqref{eq:limit:mean:omega} reduces to 
$
    2\pi\EX{\denI} KT$.
\end{remark}

\begin{figure}[t!]
    \centerline{\includegraphics[width=0.55\textwidth]{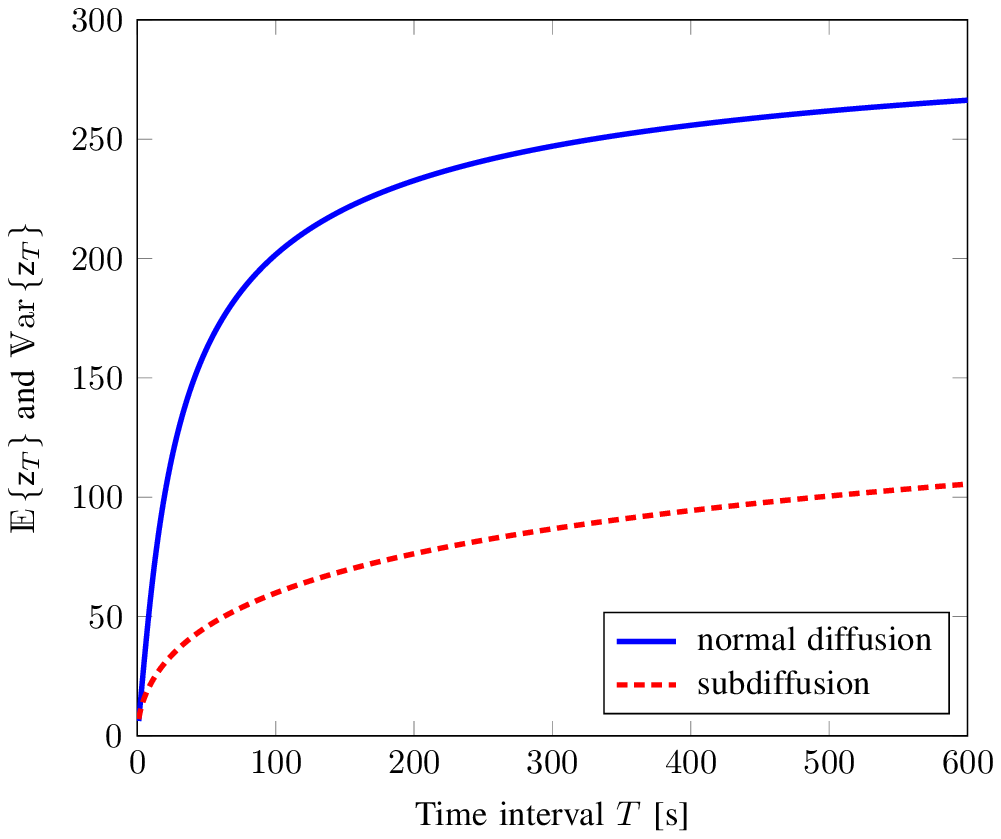}}
\caption{$\EX{\rv{z}_T}$ and $\Var{\rv{z}_T}$ as functions of the time interval $T$ in the $\left(\alpha,\beta\right)$-anomalous diffusion channel with $\EX{\rv{\lambda}}=10^{10}$~[molecules/m$^2$] and $\omega=10^{-4}$~[$\mathrm{m}$] when: i) $\left(\alpha,\beta\right)=\left(2,1\right)$ for normal diffusion; and ii) $\left(\alpha,\beta\right)=\left(2,0.5\right)$ for subdiffusion.
}
\label{fig:9}
\end{figure}

\begin{figure}[t!]
    \centerline{\includegraphics[width=0.55\textwidth]{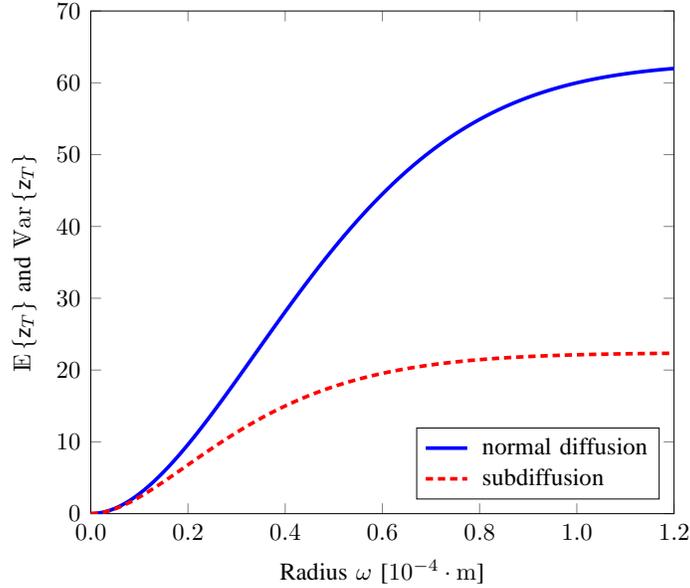}}
\caption{$\EX{\rv{z}_T}$ and $\Var{\rv{z}_T}$ as functions of the radius $\omega$ in the $\left(\alpha,\beta\right)$-anomalous diffusion channel with $\EX{\rv{\lambda}}=10^{10}$~[molecules/m$^2$] and $T=10$ when: i) $\left(\alpha,\beta\right)=\left(2,1\right)$ for normal diffusion; and ii) $\left(\alpha,\beta\right)=\left(2,0.5\right)$ for subdiffusion.
}
\label{fig:10}
\end{figure}

\begin{figure}[t!]
    \centerline{\includegraphics[width=0.55\textwidth]{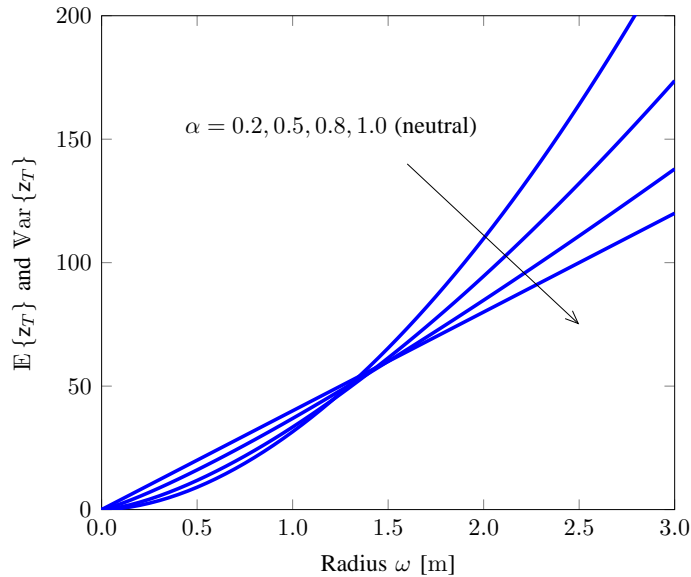}}
\caption{$\EX{\rv{z}_T}$ and $\Var{\rv{z}_T}$ as functions of the radius $\omega$ in the $\left(\alpha,1\right)$-anomalous diffusion channel with $\EX{\rv{\lambda}}=10^{10}$~[molecules/m$^2$] and $T=10$ when $\alpha=0.2,0.5,0.8$ and $1.0$ (neutral fractional diffusion)
}
\label{fig:11}
\end{figure}

\begin{example}
Fig.~\ref{fig:9} shows the mean and variance of $\rv{z}_T$ as a function of the time interval $T$ in the $\left(\alpha,\beta\right)$-anomalous diffusion channel with $\EX{\rv{\lambda}}=10^{10}$~[molecules/m$^2$] and $\omega=10^{-4}$~[$\mathrm{m}$] when $\left(\alpha,\beta\right)=\left(2,1\right)$ for normal diffusion and $\left(\alpha,\beta\right)=\left(2,0.5\right)$ for subdiffusion. As expected, $\EX{\rv{z}_T}$ and $\Var{\rv{z}_T}$ are  monotonically increasing with respect to the time interval $T$ until reaching the maximum number of interfering molecules ($314$~[molecules]). Note that $\EX{\rv{z}_T}$ and $\Var{\rv{z}_T}$ for the superdiffusion scenario have very small values in this example.
To demonstrate the behavior of $\rv{z}_T$ in a large area, we plot $\EX{\rv{z}_T}$ and $\Var{\rv{z}_T}$ as a function of the radius $\omega$ in $\left(\alpha,\beta\right)$-anomalous diffusion in Figs.~\ref{fig:10} and~\ref{fig:11}, where $\EX{\rv{\lambda}}=10^{10}$~[molecules/m$^2$] and $T=10$~[s] for normal diffusion, subdiffusion (Fig.~\ref{fig:10}) and superdiffusion (Fig.~\ref{fig:11}). We observe that $\EX{\rv{z}_T}$ and $\Var{\rv{z}_T}$ increase quickly with $\omega$ until reaching their limits in \eqref{eq:limit:mean:omega} for $\alpha=2$. In this example, the maximum mean and variance of $\rv{z}_T$  are equal to $62.83$~[molecules] for the normal diffusion and $22.42$~[molecules] for the $\left(2,0.5\right)$-anomalous diffusion, respectively. On the other hand, we see from Fig.~\ref{fig:11} that $\EX{\rv{z}_T}$ and $\Var{\rv{z}_T}$ increase nonlinearly with $\omega$ for the superdiffusion scenario except for the case of $\alpha=\beta=1$, where $\EX{\rv{z}_T}$ and $\Var{\rv{z}_T}$ increase linearly with $\omega$ (see Remark~\ref{rem:inf:time}).
\end{example}

The Poisson distribution proposed in \cite{JC:60:JAMS} can be applied when the number of interferers  goes to infinity and the probability that interfering molecules arrive at the RN during the interval $T$ tends to zero. In this paper, we consider a Gaussian distribution with mean $\mu_{T}=\EX{\rv{z}_T}$ and variance $\sigma_{T}^2=\Var{\rv{z}_T}$ to model the number of arriving interfering molecules. 
This Gaussian approximation is well fitted to the large number of interfering molecules 
under Lindeberg's condition for the central limit theorem \cite{Bil:95:Book}.

\subsection{Bit Error Rate Analysis in the Presence of Interfering Molecules}

Let
\begin{align} 
\rv{t}_\mathrm{I}^\star
&=
	\min_{\RV{x} \in \rv{\Psi}\left(\cR\right)}
	t \left(\RV{x}\right)
\end{align}
be the minimum FPT of interfering molecules, where $t\left(\RV{x}\right)$ denotes the FPT of the interfering molecule located at $\RV{x} \in \rv{\Psi}\left(\cR\right)$. Then, the arrival time in \eqref{eq:at:time} can be rewritten in the presence of interfering molecules as
\begin{align}
    \tilde{\rv{y}}_{\mathrm{tm},\ell}
    =    \rv{s}_\ell + \tilde{\rv{t}}_{\ell}
\end{align}
where $\tilde{\rv{t}}_\ell = \min \left\{\rv{t}\left(\rv{r}_\ell\right), \rv{t}_\mathrm{I}^\star\right\}$ denotes the FPT of the first arrival molecule.

\begin{theorem}   \label{thm:ber:withoutI}
Let $\rv{r}_\ell\sim \FoxV{\pM}{\pN}{\pP}{\pQ}{\pSeq_\ell}$ be the random distance from the $\ell$th nearest TN, $\omega$ be the radius of $\cR$, and $\rv{\lambda}$ be the random intensity for the Cox process $\rv{\Psi}$ of interfering molecules. Then, the BER $\IPb{\ell}$ of molecular communication between the $\ell$th nearest TN and RN with timing modulation in $\left(\alpha,\beta\right)$-anomalous diffusion for a fixed detection threshold $\gamma=\Tb/2$ is given by
\begin{align}	\label{eq:thm:berti}
    \IPb{\ell}
    &=
        \frac{1}{2}
        \left[
        1+
                \left(2\hat{P}_{\mathrm{b},\ell}-1\right)       
                \exp
                \left(
                -
       		4\pi\EX{\rv{\lambda}}
        		\frac{\left(\Tb/2\right)^{2\beta/\alpha}}{\alpha}
        		\Fox{2}{2}{4}{4}{\omega \left(\Tb/2\right)^{-\beta/\alpha};\pSeqImean}                
                \right)
         \right]
\end{align}
where $\pSeqImean$ is given in \eqref{eq:pSeqImean}.
\begin{proof}
Without knowledge of the interference distribution, the information can be decoded by the first arrival molecule. 
Since 
\begin{align} 
        &\Prob{\min\left\{\rv{t}\left(\rv{r}_\ell\right),\rv{t}_\mathrm{I}^\star \right\} \leq t | \rv{\Psi}\left(\cR\right)}
=1-\left(1-\CDF{\rv{t}\left(\rv{r}_\ell\right)}{t}\right)
        \prod_{\RV{x} \in \rv{\Psi}\left(\cR\right)}\left(1-q_{t}\left(\RV{x}\right)\right),
\end{align}
the conditional BER can be formulated as 
\begin{align} \label{eq:tm:inf:ber}
    \IPb{\ell}\left(\rv{\Psi}_\cR\right)
     &=
     \frac{1}{2}
 	\EX{
	\Prob{\tilde{\rv{y}}_{\mathrm{tm},\ell}>\left.\frac{\Tb}{2}\right|\rv{s}_\ell=0,\rv{\Psi}\left(\cR\right), \rv{r}_\ell}
        + 
        \Prob{\tilde{\rv{y}}_{\mathrm{tm},\ell}<\left.\frac{\Tb}{2}\right|\rv{s}_\ell=\frac{\Tb}{2},\rv{\Psi}\left(\cR\right), \rv{r}_\ell}
        }
    \nonumber \\
    &=    
    \frac{1}{2}
        \biggl(1
                    -
                    \EX{\CDF{\rv{t}\left(\rv{r}_\ell\right)}{\frac{\Tb}{2}}}
            \prod_{\RV{x} \in \rv{\Psi}\left(\cR\right)}\left(1-q_{\Tb/2}\left(\RV{x}\right)\right)
        \biggr).
\end{align}
Therefore, the BER $\IPb{\ell}$ is given by
\begin{align} 
\IPb{\ell}
	&=\EX{\IPb{\ell}\left(\rv{\Psi}_\cR\right)}
\nonumber \\
&=
        \frac{1}{2}
        \Bigg[1+
                \left(2\hat{P}_{\mathrm{b},\ell}-1\right)
                \mathbb{E}
                \,
                \Bigg\{
                \prod_{\RV{x} \in \rv{\Psi}\left(\cR\right)}
                \left(1-q_{\Tb/2}\left(\RV{x}\right)\right)
	\Bigg\}
        \Bigg].
\end{align}
Finally, using \eqref{eq:qti}  and the probability generating functional of the PPP \cite[Definition A.5]{HG:08:FTN}, we arrive at the desired result.
\end{proof}
\end{theorem}



Theorem~\ref{thm:ber:withoutI} shows that the existence of interfering molecules degrades the BER performance significantly. As a simple interference avoidance technique with knowledge of the interference distribution at the RN, we can improve the BER performance by using the ($\mu_T+1$)-th arriving molecule to decode the transmit information.
\begin{theorem}	\label{thm:ber:timing:withID}
Let $\omega$ be the radius of $\cR$ and $\rv{\lambda}$ be the random intensity for the Cox process $\rv{\Psi}$ of interfering molecules. Suppose that the RN decodes the information bit based on the ($n+1$)-th arriving molecule using the detection threshold $\gamma=\Tb/2$. Then, using the Gaussian approximation to $\rv{z}_{\Tb/2} \sim \RG{\mu_{\Tb/2}}{\sigma^2_{\Tb/2}}$, the optimal value of $n$ that minimizes the BER $\IPb{\ell}^\star$ of molecular communication between the $\ell$th nearest TN and the RN with timing modulation in $\left(\alpha,\beta\right)$-anomalous diffusion is equal to
$
n=\mu_{\Tb/2}
$, 
and the corresponding BER $\IPb{\ell}^\star$ is
\begin{align}	\label{eq:ber:twint}
    \IPb{\ell}^\star
    &\approx
        \frac{1}{2}
        \left[
        1+
        \left(2\hat{P}_{\mathrm{b},\ell}-1\right)        
        \left(1-2Q\left(\frac{1}{2\sigma_{\Tb/2}}\right)\right)
        \right].
\end{align}
\begin{proof}
For equiprobable bits `0' and `1', we have
\begin{align}	\label{eq:59}
    \IPb{\ell}^\star
	&=
	\frac{1}{2}
	\,
	\mathbb{E}\biggl\{
	\Prob{\left(\rv{y}_{\mathrm{tm},\ell} > \frac{\Tb}{2} \cap \rv{z}_{\Tb/2} = n\right) \bigcup \left(\rv{z}_{\Tb/2} <n \right)|\rv{s}_\ell = 0,\rv{r}_\ell}
	\nonumber \\
	&\hspace{1cm}+
	\Prob{\left(\rv{y}_{\mathrm{tm},\ell}< \frac{\Tb}{2} \cap \rv{z}_{\Tb/2}=n\right) \bigcup \left(\rv{z}_{\Tb/2} >n \right)|\rv{s}_\ell = \frac{\Tb}{2},\rv{r}_\ell}
	\biggr\}
	\nonumber \\
	&
	=
	\hat{P}_{\mathrm{b},\ell} \Prob{\rv{z}_{\Tb/2}=n}
	+
	\frac{1}{2}\left(1-\Prob{\rv{z}_{\Tb/2}=n}\right)
	\nonumber \\
	&
	\approx
	\frac{1}{2}
	\left[
	1-\Prob{n-\frac{1}{2}<\rv{z}_{\Tb/2}<n+\frac{1}{2}}
	\left(1-2\hat{P}_{\mathrm{b},\ell}\right)
	\right]	
\end{align}
where the last step follows from the \emph{continuity correction} such that a continuous distribution to approximate a discrete one. Using the Gaussian approximation $\rv{z}_{\Tb/2}\sim \RG{\mu_{\Tb/2}}{\sigma^2_{\Tb/2}}$, the minimum $\IPb{\ell}^\star$ can be obtained by setting $n=\mu_{\Tb/2}$ and we get \eqref{eq:ber:twint}. 
\end{proof}
\end{theorem}

\begin{remark} \label{rem:var:I}
Degradation of BER in Theorem~\ref{thm:ber:timing:withID} depends on the spatial variance of the number of arrival interfering molecules. As $\sigma^2_{\Tb/2} \rightarrow \infty$ (strong interference effect), $\IPb{\ell}^\star$ approaches $1/2$. On the contrary, as $\sigma^2_{\Tb/2} \rightarrow 0$ (no interference effect), we have
$\IPb{\ell}^\star\rightarrow\hat{P}_{\mathrm{b},\ell}$.

\end{remark}

\begin{figure*}[t!]
\centering
    \subfigure[~$\EX{\rv{\lambda}}=10^5$]{
        \includegraphics[width=0.55\textwidth]{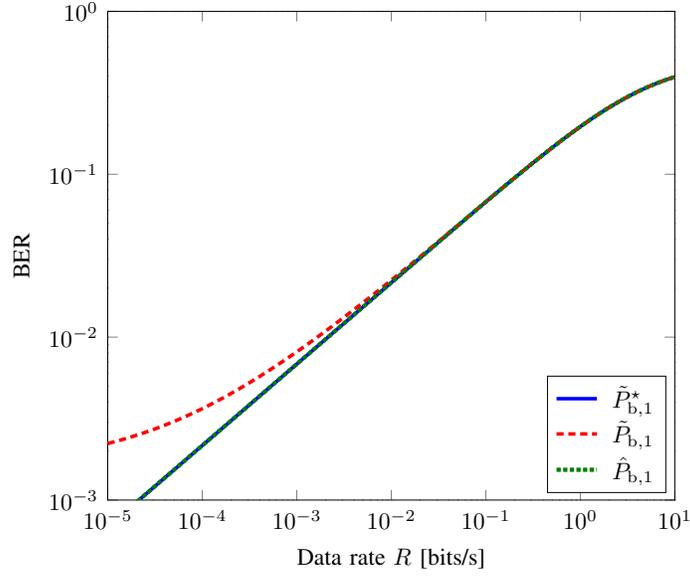}
        \label{fig:12a}
    }\hfill
    \subfigure[~$\EX{\rv{\lambda}}=10^6$]{
        \includegraphics[width=0.55\textwidth]{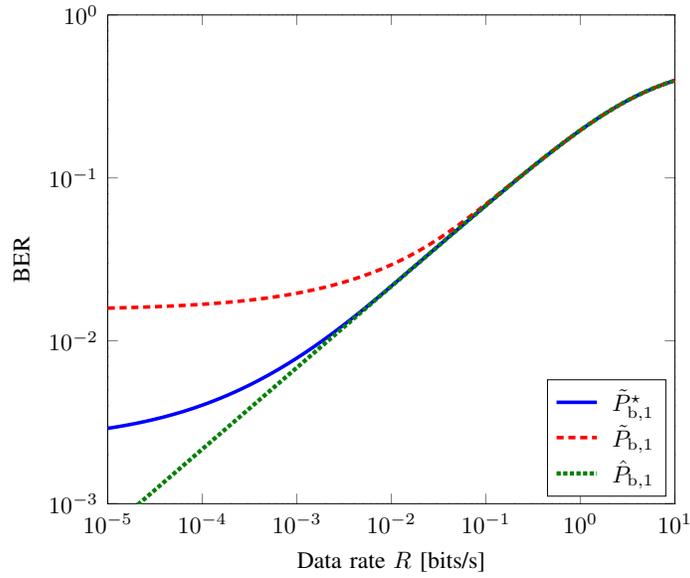}
        \label{fig:12b}
    }
\caption{BERs $\IPb{1}^\star$, $\tilde{P}_{\mathrm{b},1}$, and $\hat{P}_{\mathrm{b},1}$ as functions of the data rate $R$ for the nearest molecular communication with timing modulation in the normal diffusion channel with the Cox $\left(5,0.2\times 10^{10}\right)$-gamma field of TNs and $\omega=10^{-4}$~[$\mathrm{m}$] when (a) $\EX{\rv{\lambda}}=10^{5}$ and (b) $10^{6}$~[molecules/$\mathrm{m}^2$]. 
}
\label{fig:12}
\end{figure*}


\subsection{Numerical Examples}
Fig.~\ref{fig:12} shows the BERs $\IPb{1}^\star$, $\tilde{P}_{\mathrm{b},1}$, and $\hat{P}_{\mathrm{b},1}$ as functions of the data rate $R$ for the nearest molecular communication with timing modulation in the normal diffusion channel with the Cox $\left(5,0.2\times 10^{10}\right)$-gamma field of TNs and $\omega=10^{-4}$~[$\mathrm{m}$] when (a) $\EX{\rv{\lambda}}=10^{5}$ and (b) $10^{6}$~[molecules/$\mathrm{m}^2$]. The degradation of BER increases with the average spatial density $\EX{\rv{\lambda}}$,  as expected. As can be seen from both figures, the BER cannot achieve the low-rate slope in the presence of interfering molecules without knowledge of the interference distribution. This is because the probability that interfering molecules arrive at the RN within $\Tb/2$ also increases as the data rate decreases. On the other hand, the interference avoidance scheme with knowledge of the interference distribution can suppress the interference effect in both situations and can successfully alleviate the interference effect in low density interfering molecules nanonetworks. 




\section{Conclusions}


Using anomalous diffusion-based molecular communication channels and general forms of a spatial stochastic process, we developed the framework to characterize the $\ell$th nearest molecular communication in stochastic nanonetworks. With a versatile family of statistical distributions---i.e., $H$-variates---for the random distance between the TNs and RN in anomalous diffusion, the FPT can be formulated as again an $H$-variate in a unified fashion. Without accounting for interfering molecules, we analyzed the BER performance with timing modulation. We further determined the low-rate slope to characterize the effects of anomalous diffusion and the stochastic nature of molecules on the BER performance. In the presence of interfering molecules, we characterized the mean and variance of the number of interfering molecules arriving in a given time interval by averaging the spatial process over all  space.  It was shown that significant BER performance degradation was caused by interfering molecules with timing modulation, which can be overcome with statistical knowledge of the interfering molecules. These results are applicable for various molecular communication systems, each with unique diffusive propagation and spatial characterizations, such as relay (multihop) molecular communication systems, multiple-input-multiple-output molecular communication systems, and molecular sensor systems. For example, a relay molecular communication system is one solution to increase the molecular communication range. The performance of $\ell$th nearest molecular communication in a network answers the question of how many TNs can be reliably connected with the RN given a BER threshold in a region, or which is the optimal nearest TN for relaying this information. It is challenging to design and analyze the optimal positioning of relay nodes to  increase communication coverage in nanoscale networks with spatially distributed and moving nanomachines (nanosensors). The spatial average and ordering of the error rate achieved by the $\ell$th nearest TNs is applicable for determining routing strategies in relay and multihop molecular communication systems. It is also noteworthy that even though we considered a single fixed RN at the origin, the $\ell$th nearest molecular communication scenario can be extended to either molecular communication between two arbitrary nodes or two neighboring nodes.



\appendix[Glossary of Notation and Symbols]	\label{sec:appendix:NS}


\begin{basedescript}{\desclabelwidth{2.8cm}}
{

\item[~$\R$]

Real numbers

\item[~$\R_+$]

Nonnegative real numbers 

\item[~$\R_{++}$]

Positive real numbers  

\item[~$\Z_{+}$]

Nonnegative integers  



\item[~$\B{1}_n$] 

All-one sequence or vector of $n$ elements  

\item[~$\EX{\cdot}$] 

Expectation operator  

\item[~$\Var{\cdot}$] 

Variance operator  

\item[~$\aeq$] Asymptotically equivalent: 
$
f\left(x\right) \aeq g\left(x\right)
    ~\Leftrightarrow~
    \lim_{x \rightarrow \infty}
    \frac{f\left(x\right)}{g\left(x\right)}=1
$
 
\item[~$\doteq$]  

Asymptotically exponential equality:
$
f\left(x\right) \doteq x^y
    ~\Leftrightarrow~
    \lim_{x \rightarrow \infty}
        \frac{\log f\left(x\right)}{\log x}
    =y
$ 
 
\item[]
where $y$ is called the exponential order of $f\left(x\right)$  

\item[~$\PDF{\rv{x}}{x}$]

Probability density function of $\rv{x}$  

\item[~$\CDF{\rv{x}}{x}$] 

Cumulative distribution function of $\rv{x}$   

\item[~$\delta\left(x\right)$]

Dirac delta function  

\item[~$Q\left(\cdot\right)$] 

$Q$-function   

\item[~$\GF{\cdot}$] 

Gamma function \cite[eq.~(8.310.1)]{GR:07:Book}  

\item[~$I_x\left(a,b\right)$] 

Regularized incomplete beta function \cite[eq.~(8.392)]{GR:07:Book}

\item[~$H^{\pM,\pN}_{\pP,\pQ}{\left[\cdot\right]}$]

Fox's $H$-function \cite{JSW:15:IT}: 
\begin{flalign}
\Fox{\pM}{\pN}{\pP}{\pQ}{x;\pSeq}
&=
	\pK
	\FoxH{\pM}{\pN}{\pP}{\pQ}{\pC x}{
        \left(\pA{1},\pS{1}\right),
		\left(\pA{2},\pS{2}\right),
		\ldots,
		\left(\pA{\pP},\pS{\pP}\right)}{
		\left(\pB{1},\pT{1}\right),
		\left(\pB{2},\pT{2}\right),
		\ldots,
		\left(\pB{\pQ},\pT{\pQ}\right)}
	\nonumber\\
&=
	\pK
	\FoxH{\pM}{\pN}{\pP}{\pQ}{\pC x}
		{\left(\aV,\sV\right)}
		{\left(\bV,\tV\right)}
&
\end{flalign}
where the parameter sequence is 
$
\pSeq
=
	\left(\pK,\pC,\aV,\bV,\sV,\tV\right)
$
with 
\begin{flalign}
&
\begin{cases}
\aV=\left(\pA{1},\pA{2},\ldots,\pA{\pN},\pA{\pN+1},\pA{\pN+2}\ldots,\pA{\pP}\right) 
\\
\bV=\left(\pB{1},\pB{2},\ldots,\pB{\pM},\pB{\pM+1},\pB{\pM+2},\ldots,\pB{\pQ}\right)
\\
\sV=\left(\pS{1},\pS{2},\ldots,\pS{\pN},\pS{\pN+1},\pS{\pN+2},\ldots,\pS{\pP}\right)
\\
\tV=\left(\pT{1},\pT{2},\ldots,\pT{\pM},\pT{\pM+1},\pT{\pM+2},\ldots,\pT{\pQ}\right)
\end{cases}
&
\end{flalign}
A Mellin-Barnes type integral form of Fox's $H$-function is 
\begin{flalign}
\Fox{\pM}{\pN}{\pP}{\pQ}{x;\pSeq}
&=
	\frac{1}{2\pi \jmath}
    \int_\mathfrak{L}
    \theta\left(s\right)
    x^s
    ds,\quad x\neq 0
&
\end{flalign}
where $\mathfrak{L}$ is a suitable contour, $\jmath=\sqrt{-1}$, 
$
x^s
=
    \exp\left\{s\left(\ln\left|x\right|+\jmath \arg x\right)\right\}
$,
and 
\begin{flalign}
\theta\left(s\right)
&=
    \frac{
            \prod_{j=1}^{\pM}\GF{\pB{j}-\pT{j}s}
            \prod_{j=1}^{\pN}\GF{1-\pA{j}+\pS{j}s}
         }
         {
            \prod_{j=\pM+1}^{\pQ}\GF{1-\pB{j}+\pT{j}s}
            \prod_{j=\pN+1}^{\pP}\GF{\pA{j}-\pS{j}s}
         }
&
\end{flalign}

\item[~$\FoxHT{\pM,\pN}{\pP,\pQ}{\pSeq}{f\left(t\right)}{s}$]

$H$-transform of a function $f\left(t\right)$ with Fox's $H$-kernel of the order sequence  
\item[]
$\oSeq=\left(\pM, \pN, \pP, \pQ\right)$ and the parameter sequence $\pSeq=\left(\pK,\pC,\aV,\bV,\sV,\tV\right)$ \cite{JSW:15:IT}:  
\begin{flalign} \label{eq:Def:HT}
\FoxHT{\pM,\pN}{\pP,\pQ}{\pSeq}{f\left(t\right)}{s}
&=
	\pK
    	\int_0^\infty
        	\FoxH{\pM}{\pN}{\pP}{\pQ}
        	{\pC st}
        	{\left(\aV, \sV\right)}
        	{\left(\bV, \tV\right)}
        	f\left(t\right)
        dt,
        \quad
        s>0
&
\end{flalign}

\item[~$\FoxV{\pM}{\pN}{\pP}{\pQ}{\pSeq}$]

$H$-variate with the order sequence $\oSeq=\left(\pM,\pN,\pP,\pQ\right)$ and the parameter sequence  
\item[]
$\pSeq = \left(\pK, \pC, \aV, \bV, \sV, \tV \right)$ \cite{JSW:15:IT}: if $\rv{x} \sim \FoxV{\pM}{\pN}{\pP}{\pQ}{\pSeq}$, then %
\begin{flalign} \label{eq:Def:FV}
\PDF{\rv{x}}{x}
&=
        \pK
        \FoxH{\pM}{\pN}{\pP}{\pQ}
        {\pC x}
        {\left(\aV,\sV\right)}
        {\left(\bV,\tV\right)},
        \qquad
        x \geq 0
&
\end{flalign} 
with the set of parameters satisfying a distributional structure such that  
\item[]
$\PDF{\rv{x}}{x} \geq 0$ for all $x \in \R_+$ and $\FoxHT{\pM,\pN}{\pP,\pQ}{\pSeq}{1}{1}=1$  

\item[~$\RG{\mu}{\sigma^2}$] 

Real Gaussian distribution with mean $\mu$ and variance $\sigma^2$  

\item[~$\rayleigh{\sigma}$] Rayleigh distribution with parameter $\sigma$: 
$
\PDF{\rv{x}}{x}
=
	\frac{x}{\sigma^2}\exp\left(-\frac{x^2}{2\sigma^2}\right),
	 \quad
	 x \geq 0
$

\item[~$\GaV{a}{b}$] Gamma distribution with shape parameter $a>0$ and scale parameter $b>0$:  
\item[]
$
\PDF{\rv{x}}{x}
=
	\frac{x^{a-1}}{\GF{a}b^a}e^{-x/b},
	\quad
	x \geq 0
$

\item[~$\GGaV{a}{b}{r}$] 

Generalized gamma distribution with shape parameters $a>0$ and $b>0$ and  

\item[]
scale parameter $r>0$: 
$
\PDF{\rv{x}}{x}
=
	\frac{r x^{a r-1}}{\GF{a}b^{a r}}\,e^{-\left(x/b\right)^r},
    \quad
	x \geq 0
$

\item[~$\BP{a}{b}{r}$] Beta prime (or beta distribution of the second kind) distribution with  
\item[]
shape parameters $\alpha>0$ and $\beta>0$, and scale parameter $\gamma>0$:  
\item[]
$
\PDF{\rv{x}}{x}
=
	r^{a}\frac{\GF{a+b} x^{a-1}}{\GF{a}\GF{b}\left(1+r x\right)^{a+b}},
    \quad
	x \geq 0
$

\item[~$\Erl{n}{\lambda}$] Erlang distribution with order $n$ and hazard rate $\lambda$:
$
\PDF{\rv{x}}{x}
=
	\frac{\lambda^n x^{n-1}e^{-\lambda x}}{\left(n-1\right)!}
    ,~
	x \geq 0
$  

\item[~$\PV{\lambda}$] 

Poisson distribution with mean $\lambda$:
$
\Prob{\rv{x}=x}
=
    \frac{\lambda^x }{x!}e^{-\lambda},
    \quad
    x \in \Z_+
$  

\item[~$\Binom{n}{p}$]

Binomial distribution with mean $np$ and variance $np\left(1-p\right)$:  
\item[]
$
\Prob{\rv{x}=x}
=
    \binom{n}{x}
    p^x\left(1-p\right)^{n-x},
    \quad
    x \in \Z_+
$  

\item[~$\NB{r}{p}$] 

Negative binomial (or P\'{o}lya) distribution with mean $\frac{pr}{1-p}$ and variance $\frac{pr}{\left(1-p\right)^2}$: 
\item[] 
$
\Prob{\rv{x}=x}
=
    \frac{\GF{x+r}}{x!\GF{r}}\left(1-p\right)^r p^x,
    \quad
    x \in \Z_+
$
}
\end{basedescript}



\begin{thebibliography}{10}
\providecommand{\url}[1]{#1}
\csname url@samestyle\endcsname
\providecommand{\newblock}{\relax}
\providecommand{\bibinfo}[2]{#2}
\providecommand{\BIBentrySTDinterwordspacing}{\spaceskip=0pt\relax}
\providecommand{\BIBentryALTinterwordstretchfactor}{4}
\providecommand{\BIBentryALTinterwordspacing}{\spaceskip=\fontdimen2\font plus
\BIBentryALTinterwordstretchfactor\fontdimen3\font minus
  \fontdimen4\font\relax}
\providecommand{\BIBforeignlanguage}[2]{{%
\expandafter\ifx\csname l@#1\endcsname\relax
\typeout{** WARNING: IEEEtran.bst: No hyphenation pattern has been}%
\typeout{** loaded for the language `#1'. Using the pattern for}%
\typeout{** the default language instead.}%
\else
\language=\csname l@#1\endcsname
\fi
#2}}
\providecommand{\BIBdecl}{\relax}
\BIBdecl

\bibitem{AIM:10:CN}
L.~Atzori, A.~Iera, and G.~Morabito, ``The internet of things: {A} survey,''
  \emph{Comput. Netw.}, vol.~54, no.~15, pp. 2787--2805, Jun. 2010.

\bibitem{DF:15:NCN}
F.~Dressler and S.~Fischer, ``Connecting in-body nano communication with body
  area networks: {C}hallenges and opportunities of the internet of nano
  things,'' \emph{Nano Commun.\ Netw.}, vol.~6, no.~2, pp. 29--38, Jun. 2015.

\bibitem{APBK:15:COM}
I.~F. Akyildiz, M.~Pierobon, S.~Balasubramaniam, and Y.~Koucheryavy, ``The
  internet of bio-nano things,'' \emph{{IEEE} Commun. Mag.}, vol.~53, no.~3,
  pp. 32--40, Mar. 2015.

\bibitem{AFSFH:12:MWCOM}
I.~F. Akyildiz, F.~Fekri, R.~Sivakumar, C.~R. Forest, and B.~K. Hammer,
  ``{MoNaCo:} {F}undamentals of molecular nano-communication networks,''
  \emph{IEEE Wireless Commun. Mag.}, vol.~19, no.~5, pp. 12--28, Oct. 2012.

\bibitem{NSOMV:14:NB}
T.~Nakano, T.~Suda, Y.~Okaie, M.~J. Moore, and A.~V. Vasilakos, ``Molecular
  communication among biological nanomachines: {A} layered architecture and
  research issues,'' \emph{{IEEE} Trans.\ NanoBiosci.}, vol.~13, no.~3, pp.
  169--197, Sep. 2014.

\bibitem{AAB:12:COM}
B.~Atakan, O.~B. Akan, and S.~Balasubramaniam, ``Body area nanonetworks with
  molecular communications in nanomedicine,'' \emph{{IEEE} Commun. Mag.},
  vol.~50, no.~1, pp. 28--34, Jan. 2012.

\bibitem{CABK:15:BME}
Y.~Chahibi, I.~F. Akyildiz, S.~Balasubramaniam, and Y.~Koucheryavy, ``Molecular
  communication modeling of antibody--mediated drug delivery systems,''
  \emph{{IEEE} Trans. Biomed. Eng.}, vol.~62, no.~7, pp. 1683--1695, Jul. 2015.

\bibitem{BABK:16:NB}
A.~O. Bicen, I.~F. Akyildiz, S.~Balasubramaniam, and Y.~Koucheryavy, ``Linear
  channel modeling and error analysis for intra/inter--cellular {Ca$^{2+}$}
  molecular communication,'' \emph{{IEEE} Trans.\ NanoBiosci.}, vol.~15, no.~5,
  pp. 488--498, Jul. 2016.

\bibitem{SEA:12:IT}
K.~V. Srinivas, A.~W. Eckford, and R.~S. Adve, ``Molecular communication in
  fluid media: The additive inverse {G}aussian noise channel,'' \emph{{IEEE}
  Trans. Inf. Theory}, vol.~58, no.~7, pp. 4678--4692, Jul. 2012.

\bibitem{PA:14:COM}
M.~Pierobon and I.~F. Akyildiz, ``A statistical-physical model of interference
  in diffusion-based molecular nanonetworks,'' \emph{{IEEE} Trans. Commun.},
  vol.~62, no.~6, pp. 2085--2095, Jun. 2014.

\bibitem{LZMY:17:CL}
L.~Lin, J.~Zhang, M.~Ma, and H.~Yan, ``Time synchronization for molecular
  communication with drift,'' \emph{{IEEE} Commun. Lett.}, vol.~21, no.~3, pp.
  476--479, Mar. 2017.

\bibitem{JAJSS:16:COM}
V.~Jamali, A.~Ahmadzadeh, C.~Jardin, H.~Sticht, and R.~Schober, ``Channel
  estimation for diffusive molecular communications,'' \emph{{IEEE} Trans.
  Commun.}, vol.~64, no.~10, pp. 4238--4252, Oct. 2016.

\bibitem{GA:16:COM}
S.~Galm\'es and B.~Atakan, ``Performance analysis of diffusion-based molecular
  communications with memory,'' \emph{{IEEE} Trans. Commun.}, vol.~64, no.~9,
  pp. 3786--3793, Sep. 2016.

\bibitem{WEKN:04:B_J}
M.~Weiss, M.~Elsner, F.~Kartberg, and T.~Nilsson, ``Anomalous subdiffusion is a
  measure for cytoplasmic crowding in living cells,'' \emph{Biophys.\ J.},
  vol.~87, pp. 3518--3524, Nov. 2004.

\bibitem{SSS:97:BJ}
G.~J. Sch{\"u}tz, H.~Schindler, and T.~Schmidt, ``Single-molecule microscopy on
  model membranes reveals anomalous diffusion,'' \emph{Biophys. J.}, vol.~73,
  no.~2, pp. 1073--1080, Aug. 1997.

\bibitem{MK:00:PR}
R.~Metzler and J.~Klafter, ``The random walk's guide to anomalous diffusion:
  {A} fractional dynamics approach,'' \emph{Phys.\ Rep.}, vol. 339, no.~1, pp.
  1--77, Dec. 2000.

\bibitem{OBSTVB:06:MR}
E.~{\"O}zarslan, P.~J. Basser, T.~M. Shepherd, P.~E. Thelwall, B.~C. Vemuri,
  and S.~J. Blackband, ``Observation of anomalous diffusion in excised tissue
  by characterizing the diffusion-time dependence of the {MR} signal,''
  \emph{J.\ Mag.\ Res.}, vol. 183, no.~2, pp. 315--323, Dec. 2006.

\bibitem{MJD:09:BJ}
F.~Matth{\"a}us, M.~Jagodi{\v c}, and J.~Dobnikar, ``\textit{E. coli}
  superdiffusion and chemotaxis--search strategy, precision, and motility,''
  \emph{Biophys.\ J.}, vol.~97, no.~4, pp. 946--957, Aug. 2009.

\bibitem{Ric:26:PRSA}
L.~F. Richardson, ``Atmospheric diffusion shown on a distance-neighbour
  graph,'' \emph{Proc.\ Roy.\ Soci.\ A}, vol. 110, no. 756, pp. 709--737, Apr.
  1926.

\bibitem{PS:77:PRB}
G.~Pfister and H.~Scher, ``Time-dependent electrical transport in amorphous
  solid: {$\mathrm{AS_2Se_3}$},'' \emph{Phys.\ Rev.\ B}, vol.~15, no.~4, pp.
  2062--2083, Feb. 1977.

\bibitem{GNZ:85:PRL}
T.~Geisel, J.~Nierwetberg, and A.~Zacherl, ``Accelerated diffusion in
  {J}osephson junctions and related chaotic systems,'' \emph{Phys.\ Rev.\
  Lett.}, vol.~54, no.~7, pp. 616--619, Feb. 1985.

\bibitem{TSMD:12:PSA}
{\v Z}.~Tomovski, T.~Sandev, R.~Metzler, and J.~Dubbeldam, ``Generalized
  space--time fractional diffusion equation with composite fractional time
  derivative,'' \emph{J.\ Phys.\ A:Statis.\ Mech.\ Apps.}, vol. 391, no.~8, pp.
  2527--2542, Apr. 2012.

\bibitem{CTJS:15:CL}
T.~N. Cao, D.~P. Trinh, Y.~Jeong, and H.~Shin, ``Anomalous diffusion in
  molecular communication,'' \emph{{IEEE} Commun. Lett.}, vol.~19, no.~10, pp.
  1674--1677, Oct. 2015.

\bibitem{MMM:16:NB}
M.~U. Mahfuz, D.~Makrakis, and H.~T. Mouftah, ``Concentration-encoded
  subdiffusive molecular communication: Theory, channel characteristics, and
  optimum signal detection,'' \emph{{IEEE} Trans.\ NanoBiosci.}, vol.~15,
  no.~6, pp. 533--548, Sep. 2016.

\bibitem{MEDR:17:CL}
T.~C. Mai, M.~Egan, T.~Q. Duong, and M.~D. Renzo, ``Event detection in
  molecular communication networks with anomalous diffusion,'' \emph{{IEEE}
  Commun. Lett.}, vol.~21, no.~6, pp. 1249--1252, 2017.

\bibitem{Cal:07:Cell}
D.~E. Clapham, ``Calcium signaling,'' \emph{Cell}, vol. 131, no.~6, pp.
  1047--1058, Dec. 2007.

\bibitem{KTE:12:MCOM}
M.~S. Kuran, T.~Tugcu, and B.~O. Edis, ``Calcium signaling: Overview and
  research directions of a molecular communication paradigm,'' \emph{{IEEE}
  Wireless Commun.}, vol.~19, no.~5, pp. 20--27, Oct. 2012.

\bibitem{FKEC:14:JSAC}
N.~Farsad, N.-R. Kim, A.~W. Eckford, and C.-B. Chae, ``Channel and noise models
  for nonlinear molecular communication systems,'' \emph{{IEEE} J. Sel. Areas
  Commun.}, vol.~32, no.~12, pp. 2392--2401, Dec. 2014.

\bibitem{BBJK:14:NANO}
M.~T. Barros, S.~Balasubramaniam, B.~Jennings, and Y.~Koucheryavy,
  ``Transmission protocols for calcium--signaling-based molecular
  communications in deformable cellular tissue,'' \emph{{IEEE} Trans.
  Nanotechnol.}, vol.~13, no.~4, pp. 779--788, Jul. 2014.

\bibitem{SML:15:WCOM}
A.~Singhal, R.~K. Mallik, and B.~Lall, ``Performance analysis of amplitude
  modulation schemes for diffusion-based molecular communication,''
  \emph{{IEEE} Trans. Wireless Commun.}, vol.~14, no.~10, pp. 5681--5691, Oct.
  2015.

\bibitem{KYTA:12:NCN}
M.~S. Kuran, H.~B. Yilmaz, T.~Tugcu, and I.~F. Akyildiz, ``Interference effects
  on modulation techniques in diffusion based nanonetworks,'' \emph{Nano
  Commun.\ Netw.}, vol.~3, no.~1, pp. 65--73, Mar. 2012.

\bibitem{NCS:14:JSAC}
A.~Noel, K.~C. Cheung, and R.~Schober, ``A unifying model for external noise
  sources and {ISI} in diffusive molecular communication,'' \emph{{IEEE} J.
  Sel. Areas Commun.}, vol.~32, no.~12, pp. 2330--2343, Dec. 2014.

\bibitem{JCM:10:AEM}
S.~Jeanson, J.~Chadoeuf, M.~N. Madec, S.~Aly, J.~Floury, T.~F. Brocklehurst,
  and S.~Lortal, ``Spatial distribution of bacterial colonies in a model
  cheese,'' \emph{Appl. Environ. Microbiol}, vol.~77, no.~4, pp. 1493--1500,
  Dec. 2010.

\bibitem{DNGNE:17:MBSC}
Y.~Deng, A.~Noel, W.~Guo, A.~Nallanathan, and M.~Elkashlan, ``Analyzing large
  scale multiuser molecular communication via {3-D} stochastic geometry,''
  \emph{{IEEE} Trans. Mol. Biol. Multi-Scale Commun.}, vol.~3, no.~2, pp.
  118--133, Jun. 2017.

\bibitem{PA:11:SP}
M.~Pierobon and I.~F. Akyildiz, ``Diffusion-based noise analysis for molecular
  communications in nanonetworks,'' \emph{{IEEE} Trans. Signal Process.},
  vol.~59, no.~6, pp. 2532--2547, Jun. 2011.

\bibitem{LHNLC:16:NB}
Y.~Lu, M.~D. Higgins, A.~Noel, M.~S. Leeson, and Y.~Chen, ``The effect of two
  receivers on broadcast molecular communication systems,'' \emph{{IEEE}
  Trans.\ NanoBiosci.}, vol.~15, no.~8, pp. 891--900, Dec. 2016.

\bibitem{LHLCJ:17:MNL}
Y.~Lu, M.~D. Higgins, M.~S. Leeson, Y.~Chen, and P.~A. Jennings, ``Revised look
  at the effects of the channel model on molecular communication systems,''
  \emph{Micro \& Nano Lett.}, vol.~12, no.~2, pp. 136--139, 2017.

\bibitem{ESAW:17:CSTO}
H.~ElSawy, A.~Sultan-Salem, M.-S. Alouini, and M.~Z. Win, ``Modeling and
  analysis of cellular networks using stochastic geometry: {A} tutorial,''
  \emph{{IEEE} Commun. Surveys Tuts.}, vol.~19, no.~1, pp. 167--203, First
  Quarter 2017.

\bibitem{RQSW:09:JSAC}
A.~Rabbachin, T.~Q.~S. Quek, H.~Shin, and M.~Z. Win, ``Cognitive network
  interference,'' \emph{{IEEE} J. Sel. Areas Commun.}, vol.~29, no.~2, pp.
  480--493, Feb. 2011.

\bibitem{NJQTS:13:WCOM}
T.~M. Nguyen, Y.~Jeong, T.~Q.~S. Quek, W.~P. Tay, and H.~Shin, ``Interference
  alignment in a {P}oisson field of {MIMO} femtocells,'' \emph{{IEEE} Trans.
  Wireless Commun.}, vol.~12, no.~6, pp. 2633--2645, Jun. 2013.

\bibitem{JQKS:14:WCOM}
Y.~Jeong, T.~Q.~S. Quek, J.~S. Kwak, and H.~Shin, ``Multicasting in stochastic
  {MIMO} networks,'' \emph{{IEEE} Trans. Wireless Commun.}, vol.~13, no.~4, pp.
  1729--1741, Apr. 2014.

\bibitem{TJS:17:ACCESS}
D.~P. Trinh, Y.~Jeong, and H.~Shin, ``{MIMO} capacity in {B}inomial field
  networks,'' \emph{IEEE Access}, vol.~5, pp. 12\,545--12\,551, Jun. 2017.

\bibitem{GH:75:Book}
D.~L. Gerlough and M.~J. Huber, \emph{Traffic Flow Theory: {A}
  Monograph}.\hskip 1em plus 0.5em minus 0.4em\relax Washinton: Transportation
  Research Board National Research Council, 1975.

\bibitem{JSW:15:IT}
Y.~Jeong, H.~Shin, and M.~Z. Win, ``{$H$}-transforms for wireless
  communication,'' \emph{{IEEE} Trans. Inf. Theory}, vol.~61, no.~7, pp.
  3773--3809, Jul. 2015.

\bibitem{SKM:96:Book}
D.~Stoyan, W.~Kendall, and J.~Mecke, \emph{Stochastic Geometry and Its
  Applications}, 2nd~ed.\hskip 1em plus 0.5em minus 0.4em\relax John Wiley and
  Sons, 1996.

\bibitem{JCSW:13:JSAC}
Y.~Jeong, J.~W. Chong, H.~Shin, and M.~Z. Win, ``Intervehicle communication:
  {C}ox--{F}ox modeling,'' \emph{{IEEE} J. Sel. Areas Commun.}, vol.~31, no.~9,
  pp. 418--433, Sep. 2013.

\bibitem{Hae:05:IT}
M.~Haenggi, ``On distances in uniformly random networks,'' \emph{{IEEE} Trans.
  Inf. Theory}, vol.~51, no.~10, pp. 3584--3586, Oct. 2005.

\bibitem{MPS:05:CAM}
F.~Mainardi, G.~Pagnini, and R.~K. Saxena, ``Fox {$H$} functions in fractional
  diffusion,'' \emph{J.\ Comp.\ Appl.\ Math.}, vol. 178, pp. 321--331, 2005.

\bibitem{MLP:01:FCAA}
F.~Mainardi, Y.~Luchko, and G.~Pagnini, ``The fundamental solution of the
  space-time fractional diffusion equation,'' \emph{Fract.\ Calcul.\ Appl.\
  Anal.}, vol.~4, no.~2, pp. 153--192, 2001.

\bibitem{HF:13:RPP}
F.~Hofling and T.~Franosch, ``Anomalous transport in the crowded world of
  biology cells,'' \emph{Rep. Prog. Phys}, vol.~76, no.~3, p. 046602, Mar.
  2013.

\bibitem{NCL:04:PP}
D.~del Castillo-Negrete, B.~A. Carreras, and V.~E. Lynch, ``Fractional
  diffusion in plasma turbulence,'' \emph{Phys. Plasma}, vol.~11, no.~8, pp.
  3854--3864, Aug. 2004.

\bibitem{FSG:08:PRE}
D.~Fulger, E.~Scalas, and G.~Germano, ``Monte {C}arlo simulation of uncoupled
  continuous-time random walks yielding a stochastic solution of the space-time
  fractional diffusion equation,'' \emph{Phys.\ Rev.\ E.}, vol.~77, no.~2, p.
  021122, 2008.

\bibitem{AJS:18:COM}
A.~Ahmadzadeh, V.~Jamali, and R.~Schober, ``Stochastic channel modeling for
  diffusive mobile molecular communication systems,'' \emph{{IEEE} Trans.
  Commun.}, vol.~66, no.~12, pp. 6205--6220, Dec. 2018.

\bibitem{KLCKM:07:PRL}
T.~Koren, M.~A. Lomholt, A.~V. Chechkin, J.~Klafter, and R.~Metzler, ``Leapover
  lengths and first passage time statistics for {L\'{e}vy} flights,''
  \emph{Phys.\ Rev.\ Lett.}, vol.~99, no.~16, p. 160602, Oct. 2007.

\bibitem{Ata:13:CL}
B.~Atakan, ``Optimal transmission probability in binary molecular
  communication,'' \emph{{IEEE} Commun. Lett.}, vol.~17, no.~6, pp. 1152--1155,
  Jun. 2013.

\bibitem{MKHC:02:NAC}
D.~McMillen, N.~Kopell, J.~Hasty, and J.~J. Collins, ``Synchronizing genetic
  relaxation oscillators by intercell signaling,'' \emph{Proc. Nat. Acad. Sci.
  USA}, vol.~99, no.~2, pp. 679--684, Jan. 2002.

\bibitem{SMM:13:CL}
H.~ShahMohammadian, G.~G. Messier, and S.~Magierowski, ``Blind synchronization
  in diffusion-based molecular communication channels,'' \emph{{IEEE} Commun.
  Lett.}, vol.~17, no.~11, pp. 2156--2159, Nov. 2013.

\bibitem{JAS:17:NB}
V.~Jamali, A.~Ahmadzadeh, and R.~Schober, ``Symbol synchronization for
  diffusion-based molecular communications,'' \emph{{IEEE} Trans.\
  NanoBiosci.}, vol.~16, no.~8, pp. 873--887, Dec. 2017.

\bibitem{GAFYLEC:16:WCOM}
W.~Guo, T.~Asyhari, N.~Farsad, H.~B. Yilmaz, B.~Li, A.~Eckford, and C.-B. Chae,
  ``Molecular communications: Channel model and physical layer techniques,''
  \emph{{IEEE} Trans. Wireless Commun.}, vol.~23, no.~4, pp. 120--127, Aug.
  2016.

\bibitem{NCS:14:NB1}
A.~Noel, K.~C. Cheung, and R.~Schober, ``Improving receiver performance of
  diffusive molecular communication with enzymes,'' \emph{{IEEE} Trans.\
  NanoBiosci.}, vol.~13, no.~1, pp. 31--43, Mar. 2014.

\bibitem{KIK:15:NB}
M.~H. Kabir, S.~M.~R. Islam, and K.~S. Kwak, ``{D-MoSK} modulation in molecular
  communications,'' \emph{{IEEE} Trans.\ NanoBiosci.}, vol.~14, no.~6, pp.
  680--683, Sep. 2015.

\bibitem{MGMN:18:COM}
R.~Mosayebi, A.~Gohari, M.~Mirmohseni, and M.~Nasiri-Kenari, ``Type-based sign
  modulation and its application for {ISI} mitigation in molecular
  communication,'' \emph{{IEEE} Trans. Commun.}, vol.~66, no.~1, pp. 180--193,
  Jan. 2018.

\bibitem{NCS:14:NB}
A.~Noel, K.~C. Cheung, and R.~Schober, ``Optimal receiver design for diffusive
  molecular communication with flow and additive noise,'' \emph{{IEEE} Trans.\
  NanoBiosci.}, vol.~13, no.~3, pp. 208--222, Sep. 2014.

\bibitem{JAS:17:CL}
V.~Jamali, A.~Ahmadzadeh, and R.~Schober, ``On the design of matched filters
  for molecule counting receivers,'' \emph{{IEEE} Commun. Lett.}, vol.~21,
  no.~8, pp. 1711--1714, Aug. 2017.

\bibitem{MNES:12:SP}
M.~J. Moore, T.~Nakano, A.~Enomoto, and T.~Suda, ``Measuring distance from
  single spike feedback signals in molecular communication,'' \emph{{IEEE}
  Trans. Signal Process.}, vol.~60, no.~7, pp. 3576--3587, Jul. 2012.

\bibitem{WHL:15:CL}
X.~Wang, M.~D. Higgins, and M.~S. Leeson, ``Relay analysis in molecular
  communications with time-dependent concentration,'' \emph{{IEEE} Commun.
  Lett.}, vol.~19, no.~11, pp. 1977--1980, Nov. 2015.

\bibitem{CLY:18:NB}
G.~Chang, L.~Lin, and H.~Yan, ``Adaptive detection and {ISI} mitigation for
  mobile molecular communication,'' \emph{{IEEE} Trans.\ NanoBiosci.}, vol.~17,
  no.~1, pp. 21--35, Jan. 2018.

\bibitem{JFSG:18:COM}
V.~Jamali, N.~Farsad, R.~Schober, and A.~Goldsmith, ``Non-coherent detection
  for diffusive molecular communication systems,'' \emph{{IEEE} Trans.
  Commun.}, vol.~66, no.~6, pp. 2515--2531, Jun. 2018.

\bibitem{CMT:17:AAS}
B.~Cadre, G.~Massiot, and L.~Truquet, ``Nonparametric tests for {Cox}
  processes,'' \emph{J. Stat. Planning and Inference}, vol. 184, pp. 48--61,
  May 2017.

\bibitem{FMGCEG:19:MBSC}
N.~Farsad, Y.~Murin, W.~Guo, C.-B. Chae, A.~W. Eckford, and A.~Goldsmith,
  ``Communication system design and analysis for asynchronous molecular timing
  channels,'' \emph{{IEEE} Trans. Mol. Biol. Multi-Scale Commun.}, vol.~3,
  no.~4, pp. 239--253, Dec. 2017.

\bibitem{HG:08:FTN}
M.~Haenggi and R.~K. Ganti, ``Interference in large wireless networks,''
  \emph{Foundations and Trends in Networking}, vol.~3, no.~2, pp. 127--248,
  2008.

\bibitem{JC:60:JAMS}
J.~L.~H. Jr and L.~L. Cam, ``The poisson approximation to the poisson binomial
  distribution,'' \emph{Ann. Math. Statist.,}, vol.~31, no.~3, pp. 737--740,
  Apr. 1960.

\bibitem{Bil:95:Book}
P.~Billingsley, \emph{Probability and Measure}, 3rd~ed.\hskip 1em plus 0.5em
  minus 0.4em\relax John Wiley and Sons, 1995.

\bibitem{GR:07:Book}
I.~S. Gradshteyn and I.~M. Ryzhik, \emph{Table of Integrals, Series, and
  Products}, 7th~ed.\hskip 1em plus 0.5em minus 0.4em\relax San Diego, CA:
  Academic, 2007.

\end{thebibliography}
\end{document}